\documentclass[prl,superscriptaddress,floatfix,showpacs,aps,reprint,letterpaper,nobalancelastpage]{revtex4-1}
\usepackage{graphicx} 
\usepackage{color} 
\usepackage{transparent}
\usepackage{amsmath}
\usepackage{hyperref} 
\usepackage{amssymb}

\newcommand{\bra}[1]{\langle{#1}|}

\newcommand{\ket}[1]{|{#1}\rangle}

\begin{document}

\title {Self-Consistent Quantum Process Tomography}
\date{\today}
\author{Seth T. Merkel}
\affiliation{IBM T.J. Watson Research Center, Yorktown Heights, NY 10598, USA}
\author{Jay M. Gambetta}
\affiliation{IBM T.J. Watson Research Center, Yorktown Heights, NY 10598, USA}
\author{John A. Smolin}
\affiliation{IBM T.J. Watson Research Center, Yorktown Heights, NY 10598, USA}
\author{S. Poletto}
\affiliation{IBM T.J. Watson Research Center, Yorktown Heights, NY 10598, USA}
\author{A. D. C\'orcoles}
\affiliation{IBM T.J. Watson Research Center, Yorktown Heights, NY 10598, USA}
\author{B. R. Johnson}
\affiliation{Raytheon BBN Technologies, Cambridge, MA 02138, USA}
\author{Colm A. Ryan}
\affiliation{Raytheon BBN Technologies, Cambridge, MA 02138, USA}
\author{M. Steffen}
\affiliation{IBM T.J. Watson Research Center, Yorktown Heights, NY 10598, USA}
	  
\begin{abstract}
Quantum process tomography is a necessary tool for verifying quantum gates and diagnosing faults in architectures and gate design.  We show that the standard approach of process tomography is grossly inaccurate in the case where the states and measurement operators used to interrogate the system are generated by gates that have some systematic error, a situation all but unavoidable in any practical setting.  These errors in tomography can not be fully corrected through oversampling or by performing a larger set of experiments.  We present an alternative method for tomography to reconstruct an entire library of gates in a self-consistent manner.  The essential ingredient is to define a likelihood function that assumes nothing about the gates used for preparation and measurement.  In order to make the resulting optimization tractable we linearize about the target, a reasonable approximation when benchmarking a quantum computer as opposed to probing a black-box function.
\end{abstract}
\maketitle

\section{Introduction}

The design and control of scalable quantum architectures at the level of precision necessary for fault tolerant computation remains a considerable challenge.  There has been remarkable progress in numerous physical systems (e.g.~superconducting circuits \cite{Chow2012,dicarlo_2009,Bialczak2010}, atomic systems \cite{Haffner_multiparticle_ions_2005,gaebler2012}, etc.) to improve metrics such as the gate fidelity \cite{nielsen_gatefid_2002}, and incrementally these values are approaching the fault-tolerant threshold.  Diagnosing errors and improving designs of gates and architectures however, often requires much more information than a single scalar value. Instead we need a full characterization of a quantum process, and that necessitates quantum process tomography (QPT) \cite{chuang_blackbox_1997}.

The essential idea of QPT is as follows: prepare an initial quantum state, apply the operation we would like to reconstruct,  measure the expectation value of some observable and then repeat for different initial states and observables until it is possible to retrodict the process through some form of matrix inversion.  This scales exponentially poorly with the number of qubits, $n$, since the parameters necessary to specify a general map are $O(2^{4n})$, but generally we only need to perform QPT on small subsystems (one or two qubits) of a larger architecture and can then verify that the subsystems are isolated through other means \cite{gambetta2012}.  Additionally, due to the fact that there will be errors on the measurement outcomes, if for no reason other than finite measurement statistics, the resulting process may not be physical (i.e. a completely positive trace preserving map).  When this non-physicality arises from Gaussian noise on the measurement outcomes it can be corrected using maximum likelihood estimation or through Bayesian inference techniques \cite{James_qubitmeas_2001,Hradil2004,lvosky04,deBrugh2008,rbk2010,blume2012,smollin2012,Christandl2012}. 

In addition to stochastic measurement noise, quantum process tomography should be consistent with respect to systematic noise.  Most quantum computing architectures allow only for fixed initial states and measurement operators so that the state preparation and measurement (SPAM) described for tomography involves the application of gates that may have the same degree of error as the process being interrogated.  These errors in QPT can not be corrected by gathering more measurement statistics or enforcing physicality alone. Here we show that with standard maximum likelihood techniques these errors, and in particular coherent errors, lead to fidelity estimates from QPT that can be very poor. Even more problematic is that the ratio of the QPT to SPAM errors actually can grow as the SPAM error \emph{decreases} implying that as the gate fidelities approach the threshold it can be even harder to estimate them correctly.  SPAM errors can be accounted for in fidelity estimation by using randomized benchmarking \cite{knill_randomized_2008, Magesan2011, Magesan2012,gaebler2012,Magesan2012b} but this does not provide full tomographic information about the gates in order to diagnose and correct errors.  We observe a disagreement between benchmarking \cite{Magesan2012b} and tomography \cite{Chow2012} experiments on similar samples which suggests that SPAM errors are a current limitation that needs to be addressed.  

In this manuscript we explore techniques to compensate for SPAM errors while still obtaining a full characterization of our quantum gates.  One simple technique is to oversample, not by increasing the repetitions of a given experiment, but by increasing the number of gates used for SPAM.  We show that this technique shows no improvement for general unitary errors, however it does yield an improvement for a uniform error such as a global frame transformation.  The primary subject of this work is an alternative approach to tomography that places the SPAM gates on the same footing as the gate under investigation, and then retrodicts an entire library of unitary gates in a self-consistent manner.  The `cost' of this method in terms of the number of independent measurements and post-processing is polynomially equivalent to performing standard QPT on each gate in the library being reconstructed.  Our approach assumes very little about the gate error model, in contrast to the self-correcting state tomography proposed in \cite{james2012} or the bootstrap method in \cite{Dobrovitski2010} which otherwise compensate for the same SPAM issues.

The remainder of the manuscript proceeds as follows.  In Section \ref{sec:tomo} we give a detailed description of QPT and simulate the effects of SPAM errors for a range of error models and system parameters.  In Section \ref{sec:oversamp} we show the effects of oversampling using unitary 2-designs and in section \ref{sec:overkill} we develop self-consistent gate-set tomography. Finally, in section \ref{sec:Experiment} we apply our self-consistent tomography to experimental data obtained from two independent experiments on a system consisting of a single superconducting qubit.

\section{The trouble with tomography}\label{sec:tomo}

A general quantum process on a finite $d$-dimensional Hilbert space $\mathcal{H}_d$ ($d=2^n$ for qubit systems) is a completely positive trace-preserving (CPTP) map.  There are many ways to represent such a map, $\Lambda$, but in this manuscript we will primarily use two of them: the Choi matrix \cite{Choi1975} which is given by
\begin{equation}  
\rho_\Lambda = \frac{1}{d} \sum_{ij} E_{ij} \otimes \Lambda (E_{ij}),
\end{equation}
where $E_{ij} = \ket{i}\bra{j}$, and the Pauli transfer matrix (PTM) \cite{Chow2012}, which is 
\begin{equation}  
\mathcal({R}_\Lambda)_{ij} = \frac{1}{d} \mathrm{Tr}\left( P_i \Lambda (P_j)\right),
\end{equation}
where $P_j$ denotes the set of Pauli matrices, $\{I,X,Y,Z\}^{\otimes n}$, though one could extend this to qudit systems by considering any Hermitian basis for the $P_j$'s.  The transformation between these two representations is the linear mapping,
\begin{equation}
\begin{split}
\mathcal({R}_\Lambda)_{ij} =& \textrm{Tr}\left(\rho_\Lambda P^T_j\otimes P_k \right),\\
\rho_\Lambda =& \frac{1}{d^2} \sum_{ij} \mathcal({R}_\Lambda)_{ij} P^T_j\otimes P_k,
\end{split}
\end{equation}  
 and each representation have useful properties.  The Choi matrix is positive semidefinite if and only if $\Lambda$ is completely positive.  If we define $P_0 = I_d$, then the map is trace-preserving if and only if the first row of $\mathcal{R}_\Lambda$ is the vector $(1,0,0, \ldots)$ and is unital if the first column has the same structure.  Furthermore, in the PTM picture composition becomes matrix multiplication, $\mathcal{R}_{\Lambda_1 \circ \Lambda_2}=\mathcal{R}_{\Lambda_1} \mathcal{R}_{\Lambda_2}$, since the PTM is the standard superoperator group representation defined over the Pauli basis. 

Let us define an experiment by the triple $\{\rho_j, M_j, m_j \}$ which describes preparing the system in state $\rho_j$ measuring the operator $M_j$ and obtaining an expectation value $m_j$ according to 
\begin{equation}
m_{j} = \textrm{Tr} \left(M_j \Lambda(\rho_j) \right).\label{eq:mij}
\end{equation}
We can define a vector $\vert \rho \rangle \rangle$ whose elements are $\langle \langle j \vert \rho \rangle \rangle = {\rm Tr} (P_j \rho) / d$ as well as $\langle \langle M \vert j \rangle \rangle = {\rm Tr} (M P_j)$ (the omission of the dimensional factor in the measurement is intentional and is a consequence of bounding the values of the $\mathcal{R}$ matrix between $\pm 1$).  In this form,  
\begin{equation}
m_{j} = \langle \langle M_j \vert \mathcal{R}_\Lambda \vert \rho_j \rangle \rangle = {\rm Tr} \big(\vert M_j \rangle \rangle  \langle \langle \rho_j \vert^T \mathcal{R}_\Lambda \big).\label{eq:mijR}
\end{equation}  
This final expression is simply another inner product of the matrix $\mathcal{R}_\Lambda$ with the super-operator $\vert M_j \rangle \rangle  \langle \langle \rho_j \vert$.  For the vectorization of the superoperator $\mathcal{R}_\Lambda$ we use the notation ${\bf r}_\Lambda$ as the column major vectorized $\mathcal{R}_\Lambda$ and  ${\bf s}_{j}$ as the vectorized $\vert M_j \rangle \rangle  \langle \langle \rho_j \vert$ so that
\begin{equation}
m_{j} = {\bf s}_{j}^T {\bf r}_\Lambda.
\end{equation}     
We can now express the entire set of experiments as 
\begin{equation}
{\bf m } = S^T {\bf r}_\Lambda,
\end{equation}
where the columns of the rectangular matrix $S$ are the vectors ${\bf s}_{j}$.

The first estimate for our quantum process, ${\bf r}_{\rm bare}$, arises from linear inversion.  While $S$ is typically not square we can invert $S S^T$ or, in the case of an incomplete set of measurements, find its pseudo inverse.  This leads to the bare estimate
\begin{equation}
{\bf r}_{\rm bare} = (S S^T)^{-1} S {\bf m}.\label{eq:bare_est}
\end{equation}  
The astute reader may have noticed that at no point have we enforced that our estimate ${\bf r}_{\rm bare}$ correspond to a physical map, and we will find that for many error models it does not.  

The dominant source of error, in the tomography literature at least \cite{James_qubitmeas_2001,Hradil2004,lvosky04,deBrugh2008,rbk2010,blume2012,smollin2012,Christandl2012}, is statistical error due to finite measurement statistics.  Instead of direct access to $m_{j}$ we measure 
\begin{equation}
m_{j} =   {\bf s}_{j}^T {\bf r}_\Lambda + \sqrt{N_{j}} W_{j},
\end{equation} 
where $W_{j}$ is a Gaussian random variable with mean zero and unit variance and $N_{j}$ is the noise power.  Due to the central limit theorem we can always approximate the error model as Gaussian if we repeat the measurement sufficiently many times in which case $N_{j}$ scales as $1/(\#$ of repetitions of experiment $j$).  The transformation $m'_{j} = m_{j}/\sqrt{N_{j}}$ and ${\bf s}'_{j} = {\bf s}_{j}/\sqrt{N_{j}}$ simplifies the Gaussian likelihood function to  
\begin{equation}
\mathcal{L}( {\bf m}' \vert {\bf r} ) = \exp \left(- \frac{1}{2}\sum_{j} |m'_{j} -  {\bf s'}_{j}^T {\bf r}   |^2\right).
\end{equation}
For this error model the most rigorous estimate of the gate would be to perform some sort of Bayesian estimation where ${\bf r}_{\rm Bayes} = \int {\bf r} \mathcal{L}( {\bf m}' \vert {\bf r} ){\rm d}{\bf r}/\int \mathcal{L}( {\bf m}' \vert {\bf r} ){\rm d}{\bf r}$ but these integrals are often extremely hard to calculate, especially since there is no uniquely good measure over quantum channels $ {\rm d}{\bf r}$.  A much simpler technique is to report the maximum of the likelihood function.  If the global maximum lies inside the space of physical maps then it is simply the ${\bf r}_{\rm bare}$ from Eq.~\eqref{eq:bare_est}, and if most of the support of $\mathcal{L}( {\bf m}' \vert {\bf r} )$ is also physical then the Bayesian estimate and the maximum likelihood estimate will be be approximately the same.  In the case where the bare estimate is unphysical maximizing $\mathcal{L}( {\bf m}' \vert {\bf r} )$ over the space of physical ${\bf r}$ is a semidefinte program \cite{Boyd08} and can be solved for small instances easily with optimizers such as SeDuMi \cite{sedumi}.  This is because the likelihood function is of the form of a 2-norm distance between vectors ${\bf m'}$ and $S'^T {\bf r}$ and the constraints are that the Choi matrix is positive semidefinite.  We use the approach outlined in the supplement of \cite{Chow2012} which is an extension of the state tomography techniques in \cite{silberfarb05,riofrio2011} as well as  \cite{deBrugh2008}.  

\begin{figure}
\centering
A\\
\includegraphics[width=0.45\textwidth]{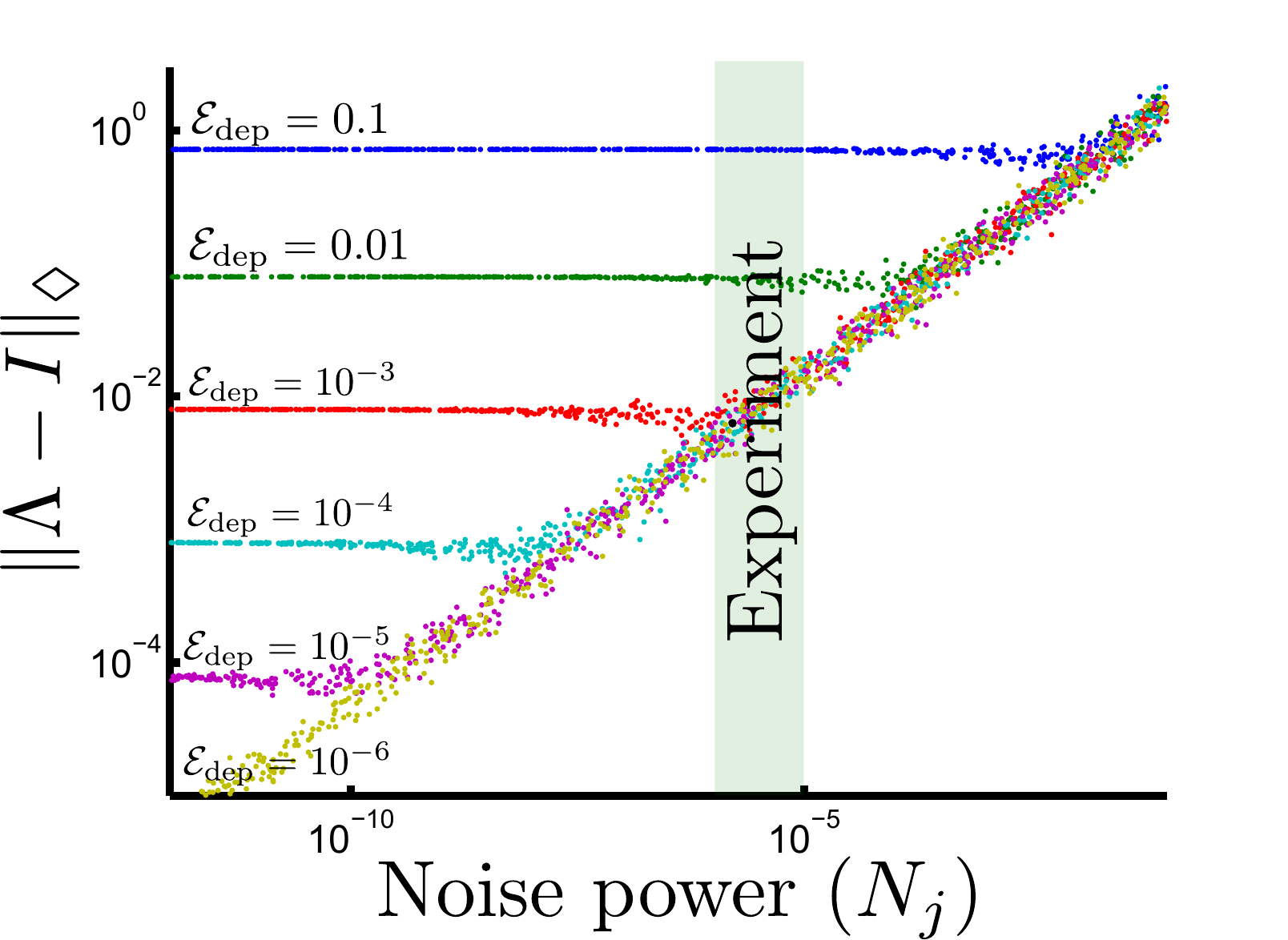}\\
B\\
\includegraphics[width=0.45\textwidth]{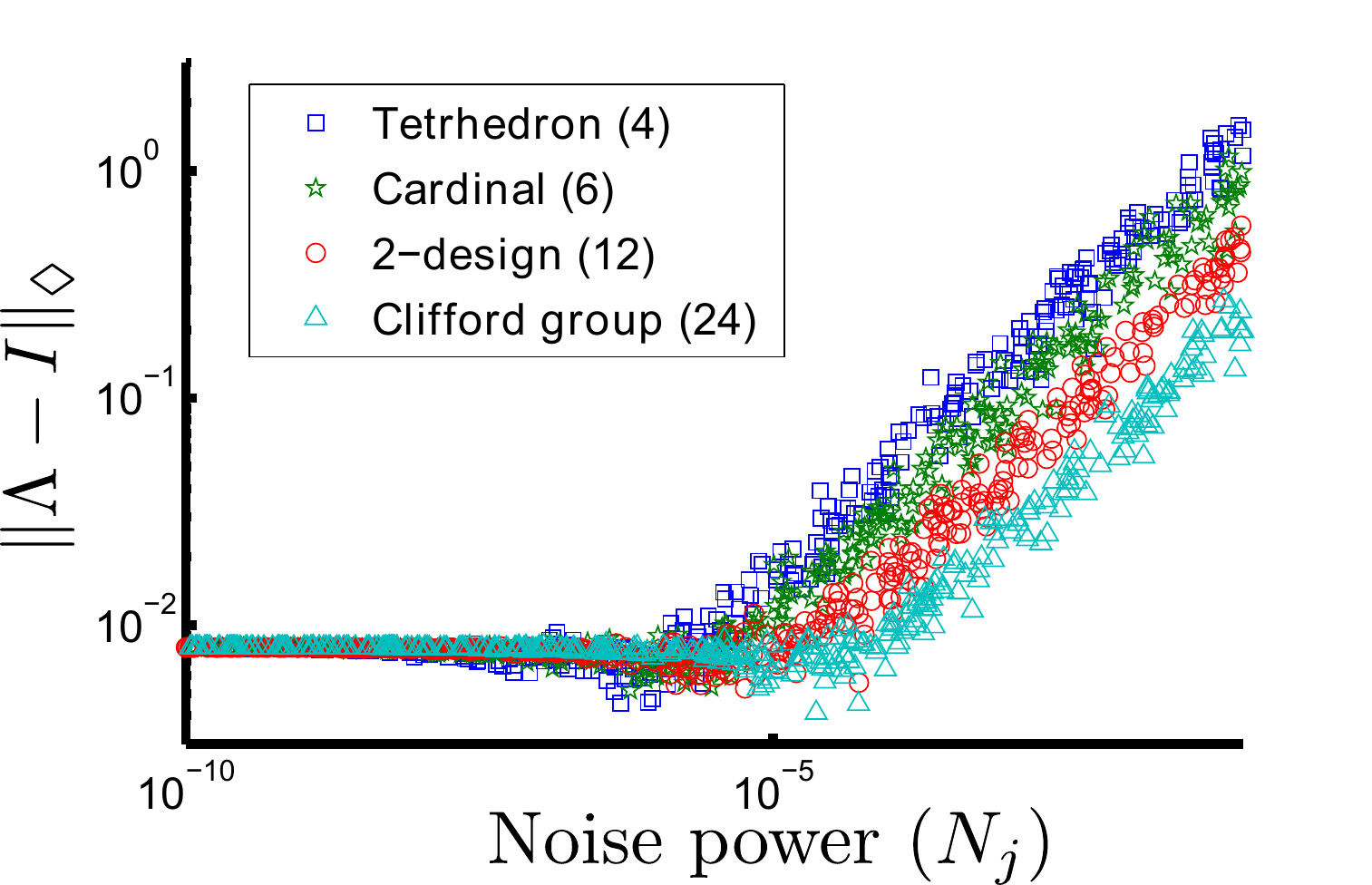}\\
\caption{(color online)  Reconstructing a perfect identity gate with imperfect tomography due to depolarizing errors on the SPAM gates as well as Gaussian random noise on the measurement outcomes.  We plot the diamond norm distance between the bare estimate Eq.~\eqref{eq:bare_est} and the identity versus the noise power.  In A) we vary the strength of the depolarizing error for a fixed gate-set with four elements corresponding to mapping the ground state to the four corners of a tetrahedron.  The experimental noise power is obtained from \cite{Chow2012}. In B) we fix the depolarizing strength at $\mathcal{E}_{\rm dep} = 10^{-3}$and vary the set of gates used for SPAM over four sets of gates of differing order.}
\label{fig:floor}
\end{figure}

\begin{figure}
\centering
\includegraphics[width=0.44\textwidth]{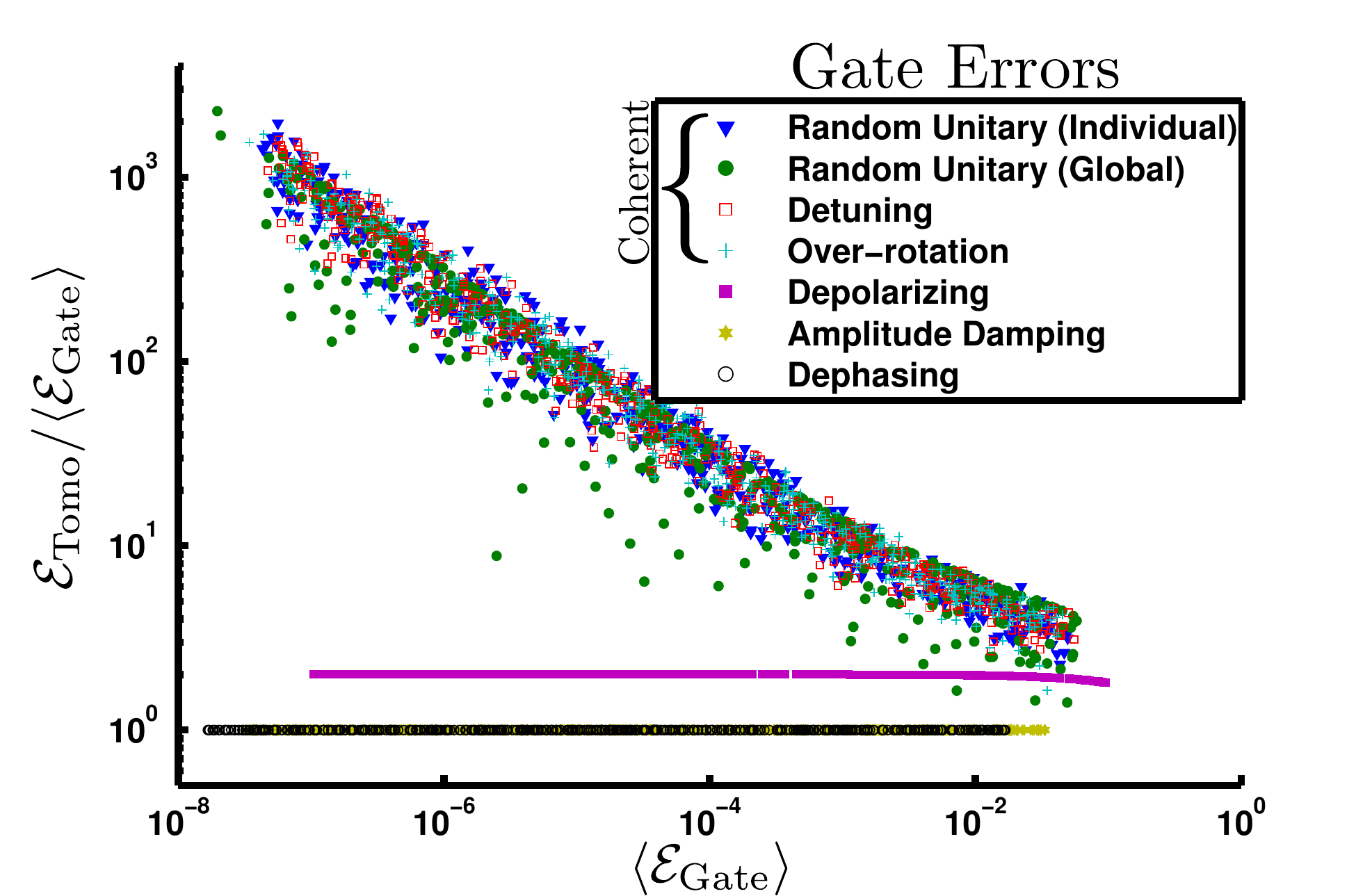}
\caption{(color online) The ratio of the error in reconstructing the identity to the average gate error plotted versus the average gate error for a single qubit using a tetrahedral gate-set and no stochastic measurement noise.  The estimated operation is the `closest' physical map to the bare reconstruction.  Seven different error models are compared, four coherent errors (different random unitary maps applied to each gate, a single random unitary map applied to each gate, a detuning error and over-rotation) as well three incoherent processes (amplitude damping, dephasing and depolarizing errors).    }
\label{fig:error_types}
\end{figure}

While there are issues regarding maximum likelihood estimation for QPT, there is a deeper problem that affects any reconstruction method that utilizes the likelihood function.  To calculate the likelihood function we require ${\bf s}_{j}$ which in turn implies we have a complete characterization of $\rho_j$ and $M_j$.  In any experimental implementation of tomography there will be SPAM errors and in many situations the magnitude of the SPAM errors is the same order as the size of the gate errors we are trying to estimate.  If the SPAM errors are stochastic, then the situation is reducible to the previous case by effectively treating the SPAM errors as additional sampling noise on the measurement.  When the errors are systematic we may write $\vert \rho_j \rangle \rangle \rightarrow \mathcal{R}_{\mathcal{E}_j}\vert \rho_j \rangle$, $\vert M_j \rangle \rangle \rightarrow \mathcal{R}_{\mathcal{F}_j}\vert M_j \rangle$  and therefore ${\bf s}_{j} \rightarrow \mathcal{R}_{\mathcal{E}_j} \otimes \mathcal{R}_{\mathcal{F}_j}^T {\bf s}_{j}$.  

For the remainder of this manuscript we will consider a slightly less general form of tomography that is applicable to many experimental implementations: a fixed initial state $\vert \rho_0 \rangle \rangle $ and measurement operator $\langle \langle M_0 \vert$ and a library of gates $G = \{ \mathcal{R}_1,\mathcal{R}_2, \ldots \mathcal{R}_N \} $.  In this picture we describe experiments according to the convention
\begin{equation}
m_{ij} = \langle \langle M_0 \vert \mathcal{R}_j \mathcal{R}_\Lambda \mathcal{R}_i\vert \rho_0 \rangle \rangle = \langle \langle M_j \vert\mathcal{R}_\Lambda\vert \rho_i \rangle \rangle.
\end{equation}
Systematic errors are described $G^{\rm (err)} = \{ \mathcal{R}_{\mathcal{E}_1} \mathcal{R}_1, \mathcal{R}_{\mathcal{E}_2}\mathcal{R}_2, \ldots  \mathcal{R}_{\mathcal{E}_N}\mathcal{R}_N \} $, under the assumption that the initial gate-set was composed of unitary maps.

In Fig.~\ref{fig:floor} we simulate the effects of both systematic and stochastic measurement noise in the case where the measured gate is a perfect identity gate.  This test is a good primitive for both theory and experiment since the identity is the one gate that should be perfectly implementable in any experiment by immediately performing measurement after state preparation (i.e.~doing nothing for no time).  In this figure the systematic SPAM error comes from a depolarizing channel of varying strengths $\mathcal{E}_{\rm dep}$.  We do not impose the CPTP constraint on the outcome and therefore measure the difference between the reconstructed gate and the identity by a diamond norm distance \cite{kitaev02}, which is calculated through the semidefinite programming technique in \cite{watrous2009}.  In this simulation the gates map the $\vert 0 \rangle$ state to the four points on a tetrahedron which is essentially the most symmetric, minimal set of gates.  We observe that for large noise powers the error in the estimate decreases exponentially with respect to decreasing noise power (and thus increasing repetitions) until it hits a floor determined by the systematic error in the interrogating gates.  

In Fig.~\ref{fig:floor}B we look at a similar plot for different libraries of unitary SPAM gates.  The first two gate-sets are defined in terms of the set of states to which they map the qubit ground state $\ket{0}$: a tetrahedron or the six cardinal directions on the Bloch sphere.  The second two groups are unitary 2-designs: a twelve element subgroup of the Clifford group and the entire 24 element Clifford group.  Both of these groups correspond to the solid rotations of a cube, with the twelve element subgroup consisting of $180^\circ$ rotations about the faces and $120^\circ$ rotations about the corners of the cube.  Generically, increasing the number of gates and maximizing the entropy of $S S^T$ speeds convergence to the estimation floor, but the actual limit remains unchanged. 

By introducing systematic errors into our interrogating gate-set the maximum of the original estimator Eq.~\eqref{eq:mij} does not yield a faithful reconstruction of the unidentified gate. In fact, the resulting estimate can be highly non-physical in the absence of any stochastic noise.  Semidefinite programming techniques constrain the estimator to the space of physical maps using the covariance matrix of the Gaussian likelihood function as a metric which we use to minimize the distance between a physical state and the unphysical estimate.  However, if this likelihood estimator is incorrect in the absence of any stochastic noise then there is no reason to trust the covariance matrix metric over any other metric on operator space (e.g.~the flat metric).  Minimizing the distance over the flat metric can be preformed using the methods in \cite{smollin2012}, which was proven to be optimal.  Since there is no gain in using the covariance method for our problem we use a flat metric in the following simulations due to its ease of calculation.

In Fig.~\ref{fig:error_types} we look at the error in reconstructing the identity, in the absence of stochastic measurement errors, for different models of the systematic errors.  We find the closest physical state to the estimate under a flat metric.  Since the two gates under comparison are physical we can compare them with the gate fidelity, $F$, or more precisely the gate error, $1-F$.  It is informative to plot the ratio of the reconstruction error to the average SPAM error on the individual gates, versus the average SPAM errors.  There is a stark difference between coherent errors (over-rotation, detuning errors etc.)and incoherent error models (T1 or T2 processes for example).  In the incoherent case the error in estimating the identity is proportional to the error on the individual gates while for the coherent model the ratio of the errors grows exponentially poorly as the gate error decreases.

The dependence of the tomography error on the gate errors has dramatic implications for benchmarking quantum gates with QPT.  A quantum computing architecture viable for fault-tolerant error correction will require gates with error rates in the range of $10^{-3}$ to $10^{-5}$.  From  Fig.~\ref{fig:error_types} we see that the \emph{measured} error rates for a fault-tolerant gate-set will be more like $10^{-2}$ to $10^{-3}$ if the dominant source of errors are coherent.  In fact, measured error rates of $10^{-5}$ are not even present in Fig.~\ref{fig:error_types} and correspond to physical gate error rates smaller than $10^{-8}$.  In many cases the dominant source of error is $T_1$ or $T_2$ type errors, in which case this effect will be minimal, however, as coherence times increase coherent errors will very likely become a bottleneck.  From Fig.~\ref{fig:floor} we see that for the typical noise power in current experiments we expect SPAM errors to dominate the estimate if the gate error is less than about a percent, which is well above the error due solely to T1 or T2.

\section{QPT with Oversampling}\label{sec:oversamp}

From the previous section one can observe that to eliminate fully stochastic errors it is sufficient to oversample, repeating experiments many times in order to accurately measure expectation values.  Oversampling is also an intuitive approach to dealing with SPAM errors if instead of increasing repetitions we alternatively increase the number of independent experiments (e.g. by increasing the size of the gate library $G$).  In this section we show that this method does \emph{not} compensate for general SPAM errors, however when the gate library is a unitary $2$-design some uniform SPAM errors are removed.

\begin{figure*}
\centering
\includegraphics[width=0.9\textwidth]{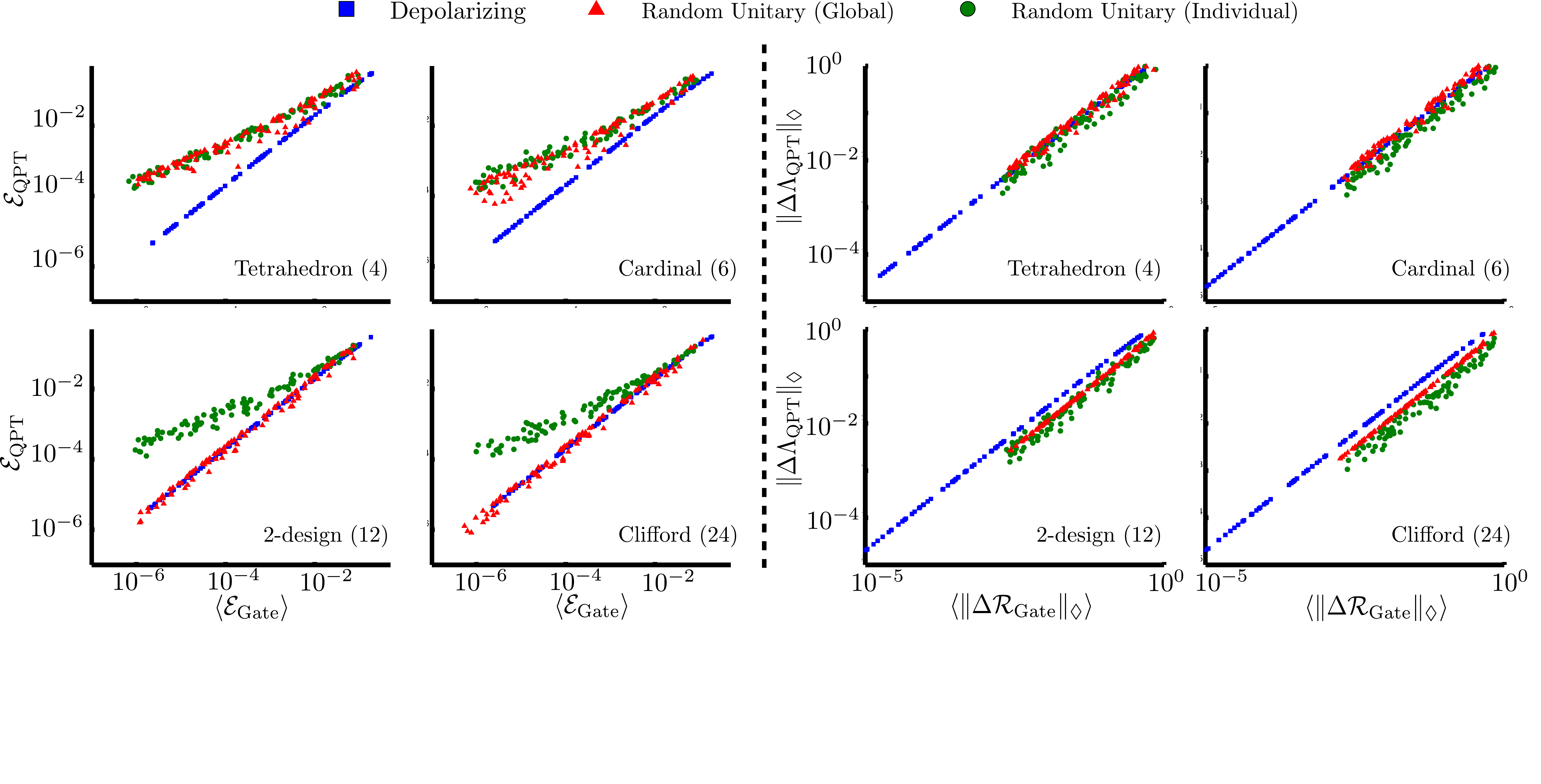}
\caption{(color online) Simulated error in reconstructing the identity versus average SPAM error in terms of the gate fidelity (left) and the diamond norm distance (right).  We look at four gate-sets and 3 types of SPAM error.}
\label{fig:overhurt}
\end{figure*}

Starting from Eq.~\eqref{eq:mijR} we can derive the bare estimate in a slightly different manner that will be more conducive to the following discussion.  First, note that we can multiply both sides by $\vert M_j \rangle \rangle  \langle \langle \rho_i \vert$ and sum over $i$ and $j$ to obtain
\begin{equation} 
\begin{split}
\sum_{ij} m_{ij}\vert M_j \rangle \rangle  \langle \langle \rho_i \vert =& \left(\sum_{j} \vert M_j \rangle \rangle    \langle \langle M_j \vert \right) \mathcal{R}_\Lambda \\
&\times\left(\sum_i \vert \rho_i \rangle \rangle \langle \langle \rho_i \vert \right).
\end{split}
\end{equation} 
In this case the bare estimate is 
\begin{equation} 
\begin{split}
\mathcal{R}_\Lambda^{({\rm est})} =& \left(\sum_{j'} \vert M_{j'} \rangle \rangle    \langle \langle M_{j'} \vert \right)^{-1} \sum_{ij} m_{ij}\vert M_j \rangle \rangle  \langle \langle \rho_i \vert\\
&\times  \left(\sum_{i'} \vert \rho_{i'} \rangle \rangle \langle \langle \rho_{i'} \vert \right)^{-1}.
\end{split}
\end{equation} 
If there are SPAM errors the term $m_{ij}$ has the form
\begin{equation}
m_{ij} =   \langle \langle M_0 \vert \mathcal{R}_{\mathcal{E}_j}\mathcal{R}_j \mathcal{R}_\Lambda \mathcal{R}_{\mathcal{E}_i} \mathcal{R}_i \vert \rho_0 \rangle \rangle
\end{equation}
which leads to,
\begin{equation} 
\begin{split}
\mathcal{R}_\Lambda^{({\rm est})} =& \left(\sum_{j'} \vert M_{j'} \rangle \rangle    \langle \langle M_{j'} \vert \right)^{-1}  \sum_{j} \mathcal{R}_j^T \vert M_0 \rangle \rangle\langle \langle M_0 \vert   \mathcal{R}_{\mathcal{E}_{j}}\mathcal{R}_{j}
\\
&\times   \mathcal{R}_\Lambda   \sum_{i} \mathcal{R}_{\mathcal{E}_{i}} \mathcal{R}_{i} \vert \rho_{0} \rangle \rangle \langle \langle \rho_{0} \vert \mathcal{R}_{i}^T \left(\sum_{i'} \vert \rho_{i'} \rangle \rangle \langle \langle \rho_{i'} \vert \right)^{-1}.
\end{split}\label{eq:full_overhurt}
\end{equation}
 
There are some very interesting consequences of the final expression which we shall examine in a few special cases.  If the SPAM error is independent of the gate, then due to our error conventions we get a cancellation of the final term in Eq.~\eqref{eq:full_overhurt} yielding,
\begin{equation} 
\begin{split}
\mathcal{R}_\Lambda^{({\rm est})} =& \left(\sum_{j} \vert M_j \rangle \rangle    \langle \langle M_j \vert \right)^{-1}\left(  \sum_{j'} \vert M_{j'} \rangle \rangle\langle \langle M_0 \vert   \mathcal{R}_{\mathcal{E}}\mathcal{R}_{j'} \right)
\\
&\times   \mathcal{R}_\Lambda  \mathcal{R}_{\mathcal{E}}.
\end{split}\label{eq:overhurt_proof}
\end{equation}  
We could derive similar expressions if the constant error occurred before the gate or in some combination of pre and post-gate error.  If in addition the error commutes with all of the gates in the library (as is the case with depolarizing errors) the result is, for example
\begin{equation}
\mathcal{R}_\Lambda^{({\rm est})} =   \mathcal{R}_{\rm dep} \mathcal{R}_\Lambda  \mathcal{R}_{\rm dep},
\end{equation}
independent of the set of gates used for tomography.  This form of the reconstructed map lends some intuition to Fig.~\ref{fig:error_types}, where the depolarizing error and other incoherent errors lead to a tomography error that had polynomial scaling with respect to the error on the gate-set.  Qualitatively, errors of this form `commute' with the reconstruction procedure.    

If the error does not commute with all of the gates but the library forms a unitary 2-design \cite{Dankert2009ca} we can also simplify equation Eq.~\eqref{eq:overhurt_proof}.  We start by independently analyzing the two quantities in parenthesis in Eq.~\eqref{eq:overhurt_proof}.  These sums are both twirls over a unitary 2-design, $\mathcal{W}$, which have the following generic form
\begin{equation}
\mathcal{W}(A) = \vert I \rangle \rangle \langle \langle I \vert  \langle \langle I \vert  A\vert I \rangle \rangle  +\left(  \mathbb{I} -   \vert I \rangle \rangle \langle \langle I \vert\right) \frac{{\rm Tr} (A) -\langle \langle I \vert  A\vert I \rangle \rangle  }{d^2-1},
\end{equation}
\cite{gambetta2012}.  Therefore, the first factor is given by
\begin{equation}
\begin{split}
\mathcal{W}&(\vert M_0 \rangle \rangle \langle \langle M_0 \vert) =\vert I \rangle \rangle \langle \langle I \vert  \langle \langle I  \vert M_0 \rangle \rangle^2  \\
 +& \left(  \mathbb{I}-   \vert I \rangle \rangle \langle \langle I \vert\right) \frac{\langle \langle M_0\vert M_0 \rangle \rangle -\langle \langle I \vert  M_0 \rangle \rangle^2 }{d^2-1}.
\end{split}
\end{equation}
and the second by
\begin{equation}
\begin{split}
\mathcal{W}&(\vert M_0 \rangle \rangle \langle \langle M_0 \vert \mathcal{R}_{\mathcal{E}}) =\vert I \rangle \rangle \langle \langle I \vert  \langle \langle I \vert  M_0 \rangle \rangle\langle \langle M_0 \vert \mathcal{R}_{\mathcal{E}} \vert I \rangle \rangle  \\
 +& \left(  \mathbb{I}-   \vert I \rangle \rangle \langle \langle I \vert\right) \frac{\langle \langle M_0 \vert \mathcal{R}_{\mathcal{E}} \vert M_0 \rangle \rangle -\langle \langle I \vert  M_0 \rangle \rangle\langle \langle M_0 \vert \mathcal{R}_{\mathcal{E}} \vert I \rangle \rangle }{d^2-1}.
\end{split}
\end{equation}
Both of these are proportional to depolarizing channels which generically have the form $\vert I \rangle \rangle \langle \langle I \vert + \epsilon (\mathbb{I} - \vert I \rangle \rangle \langle \langle I \vert )$.  The product $\mathcal{W}(\vert M_0 \rangle \rangle \langle \langle M_0 \vert)^{-1} \mathcal{W}(\vert M_0 \rangle \rangle \langle \langle M_0 \mathcal{R}_{\mathcal{E}} \vert)$ is therefore proportional to a depolarizing channel with a strength
\begin{equation}
\epsilon = \frac{\frac{\langle \langle M_0 \vert \mathcal{R}_{\mathcal{E}} \vert M_0 \rangle \rangle}{\langle \langle M_0 \vert \mathcal{R}_{\mathcal{E}} \vert I \rangle \rangle \langle \langle I \vert  M_0 \rangle \rangle}-1}{\frac{\langle \langle M_0\vert M_0 \rangle \rangle}{\langle \langle I \vert  M_0 \rangle \rangle^2} -1},
\end{equation}
and proportionality constant
\begin{equation}
\alpha = \frac{\langle \langle M_0 \vert \mathcal{R}_{\mathcal{E}} \vert I \rangle \rangle}{\langle \langle I \vert  M_0 \rangle \rangle}.
\end{equation}
The bare estimate for a constant error on a gate-set composed of a unitary 2-design is thus given by
\begin{equation}
\mathcal{R}_\Lambda^{({\rm est})} =   \alpha \mathcal{R}^{\rm (dep)}_{\epsilon} \mathcal{R}_\Lambda  \mathcal{R}_{\mathcal{E}}.
\end{equation}

The previous theoretical analysis suggests that there may be a fundamental difference when the gate library is a unitary 2-design, and this is confirmed in Fig.~\ref{fig:overhurt}.  When we measure the errors in terms of the gate fidelity we see a clear difference with respect to the unitary 2-designs.  In that case a global unitary error has the same effect on the fidelity as a depolarizing error, as expected from the previous argument.  QPT errors resulting from independent unitaries on each of the gates remains qualitatively the same for all four sets.  What we describe as a global unitary error is a strange object and should not be confused for control errors such as over rotation or detuning errors, but is instead more akin to a global frame misalignment.  

Interestingly, the errors look very different when measured by the diamond norm (the right and left plots of Fig.~\ref{fig:overhurt}). The diamond norm distance of the reconstruction scales polynomially with the gate-set diamond norm error for all forms of the SPAM error and gate-sets.  The fact that the rate of the depolarizing error spans twice as large a domain as the coherent errors in the diamond norm is also curious.  These two metrics clearly yield very different results and it is an important question as to which is appropriate for an experiment to report.  For this reason, we consider both for the remainder of this manuscript.

\section{Self-Consistent Tomography}\label{sec:overkill}

\begin{figure*}
\centering
\includegraphics[width=0.9\textwidth]{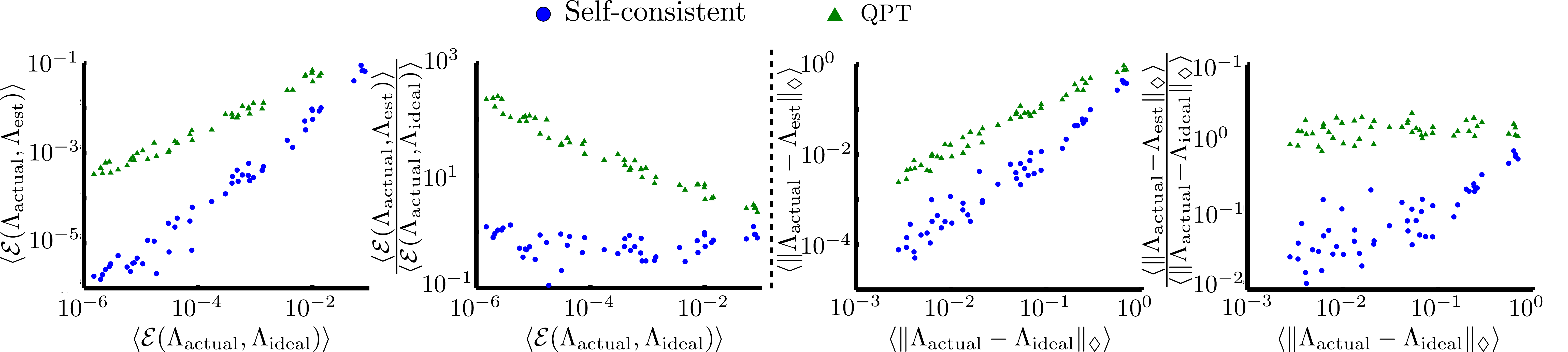}
\caption{(color online)  A comparison of standard QPT and the self-consistent method presented in this manuscript for individual random unitary errors  for the set of gates mapping to the six cardinal directions.  We plot both the diamond norm and fidelity error between the estimates and the actual gates (averaged over the set of gates) plotted versus the equivalent distance of the actual set from the ideal gates.  We show the magnitude of the tomography error and the ratio of the error with respect to the actual gate error for both cases.}
\label{fig:overkill}
\end{figure*}

\begin{figure}
\centering
\includegraphics[width=0.42\textwidth]{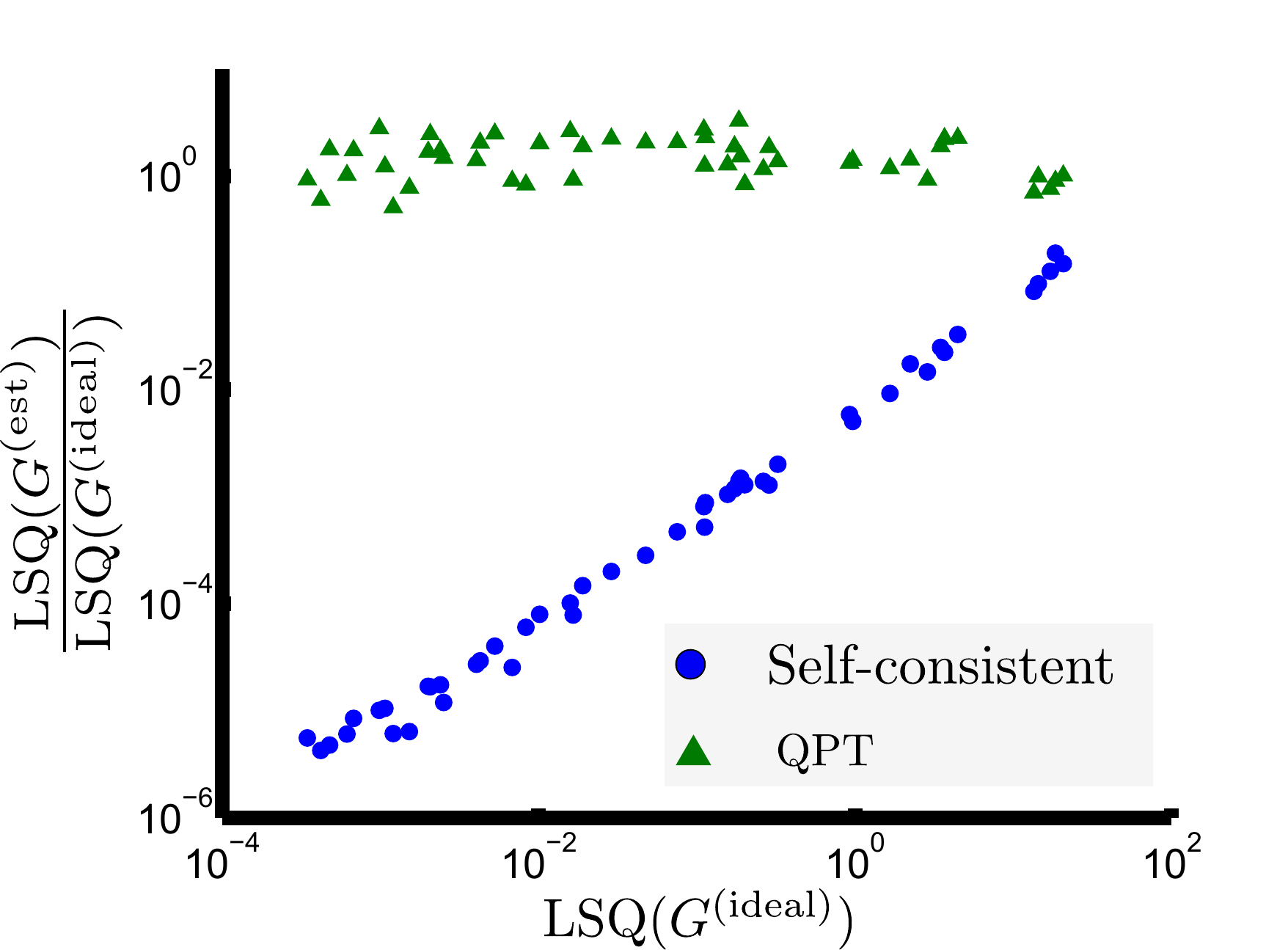}
\caption{(color online) A comparison of least-square estimator (negative log-likelihood function) for standard QPT and the linearized likelihood estimation method with SPAM errors are generated by random unitary maps of each gate. On the x-axis is the estimator of the ideal gate-set in the absence of errors with respect to the measurement outcomes and on the $y$-axis the ratio of the estimated gate-set's log likelihood to the ideal.}
\label{fig:likely}
\end{figure}
We have thus far failed at tomography in this manuscript for the simple reason that our methodology is based on the likelihood function, and we have used the wrong one.  A proper likelihood function should incorporate the SPAM errors, but characterizing the SPAM errors themselves requires QPT and therefore a likelihood function.  This is an impasse for standard QPT methods.  The alternative we present in this section is to assume nothing about the SPAM errors (though later we will bound their magnitude) and instead estimate an entire library of gates in a self-consistent manner.  

Let us write the set of experimental gates as $G^{(\rm exp)} = \{ \mathcal{R}_{\mathcal{E}_1} \mathcal{R}_1 , \mathcal{R}_{\mathcal{E}_2} \mathcal{R}_2,\ldots  \mathcal{R}_{\mathcal{E}_N} \mathcal{R}_N\}$ which is a faulty version of the library $G^{(\rm ideal)} = \{  \mathcal{R}_1 , \mathcal{R}_2,\ldots   \mathcal{R}_N\}$.  The free parameters of the this model are the error terms $\mathcal{R}_{\mathcal{E}_1}$ while the ideal gates are fixed.  We begin by performing a series of $N^3$ experiments of the form 
\begin{equation}
m_{ijk} =  \langle \langle M_0 \vert \mathcal{R}_{\mathcal{E}_k}\mathcal{R}_k \mathcal{R}_{\mathcal{E}_j}\mathcal{R}_j \mathcal{R}_{\mathcal{E}_i}\mathcal{R}_i \vert \rho_0 \rangle \rangle.
\end{equation}
which in essence consists of the tomography data necessary to reconstruct each element of $G^{(\rm exp)}$,

The likelihood of a trial set of gates $\tilde{G} = \{ \tilde{\mathcal{R}}_{\mathcal{E}_1} \mathcal{R}_1 , \tilde{\mathcal{R}}_{\mathcal{E}_2} \mathcal{R}_2,\ldots  \tilde{\mathcal{R}}_{\mathcal{E}_N} \mathcal{R}_N\}$ is now given by 
\begin{equation}
\begin{split}
\mathcal{L}(\tilde{G}) =& \exp \bigg( -\sum_{i,j,k} \big|m_{ijk} \\
&-   \langle \langle M_0 \vert \tilde{\mathcal{R}}_{\mathcal{E}_k}\mathcal{R}_k \tilde{\mathcal{R}}_{\mathcal{E}_j}\mathcal{R}_j \tilde{\mathcal{R}}_{\mathcal{E}_i}\mathcal{R}_i \vert \rho_0 \rangle \rangle \big|^2   \bigg).
\end{split}\label{eq:gen_likely}
\end{equation} 
The constraints such that the gate library is physical is that each of the maps $\tilde{\mathcal{R}}_{\mathcal{E}_j}$ are completely positive and trace preserving.  Finding a maximum likelihood estimate is equivalent to minimizing the least square estimator,
\begin{equation}
{\rm LSQ}(G) = \sum_{i,j,k} \bigg| m_{ijk} -   \langle \langle M_0 \vert \tilde{\mathcal{R}}_{\mathcal{E}_k}\mathcal{R}_k \tilde{\mathcal{R}}_{\mathcal{E}_j}\mathcal{R}_j \tilde{\mathcal{R}}_{\mathcal{E}_i}\mathcal{R}_i \vert \rho_0 \rangle \rangle \bigg|^2,
\end{equation}
but is highly non-trivial due to the fact that this function is $6^{\rm th}$ order in the gate-set.  

We can reduce the order of the estimator by linearizing the evolution about the ideal set of gates.  This is a reasonable assumption if we are testing components of a faulty quantum computer as opposed to probing some unknown physics or interaction.  In this situation we allow  $\tilde{\mathcal{R}}_{\mathcal{E}_j} = \mathcal{R}_{I}+\tilde{\mathcal{E}}_j$ where $\tilde{\mathcal{E}}_j$ is not necessarily physical but is small, $\|\tilde{\mathcal{E}}_j \| \ll 1$.  Expanding the evolution to first order in $\tilde{\mathcal{E}}_j$ we have
\begin{equation}
\begin{split}
\tilde{m}_{ijk}  =&   \langle \langle M_0 \vert\mathcal{R}_k \mathcal{R}_j \mathcal{R}_i +  \tilde{\mathcal{E}}_k\mathcal{R}_k \mathcal{R}_j \mathcal{R}_i \\
 &+   \mathcal{R}_k \tilde{\mathcal{E}}_j \mathcal{R}_j \mathcal{R}_i +    \mathcal{R}_k \mathcal{R}_j \tilde{\mathcal{E}}_i\mathcal{R}_i \vert \rho_0 \rangle \rangle,
\end{split}
\end{equation}
or by transforming back to the physical $ \tilde{\mathcal{R}}_{\mathcal{E}_j}$
\begin{equation}
\begin{split}
\tilde{m}_{ijk}  =&  \langle \langle M_0 \vert -2 \mathcal{R}_k \mathcal{R}_j \mathcal{R}_i  +   \tilde{\mathcal{R}}_{\mathcal{E}_k}\mathcal{R}_k \mathcal{R}_j \mathcal{R}_i  \\
 &+  \mathcal{R}_k \tilde{\mathcal{R}}_{\mathcal{E}_j}\mathcal{R}_j \mathcal{R}_i  +   \mathcal{R}_k \mathcal{R}_j \tilde{\mathcal{R}}_{\mathcal{E}_i}\mathcal{R}_i \vert \rho_0 \rangle \rangle.
\end{split}
\end{equation}
Therefore self-consistent reconstruction of a set of gates can be described as the following least square estimation 
\begin{equation}
\begin{split}
\min_{\tilde{\mathcal{R}}_{\mathcal{E}_1},\tilde{\mathcal{R}}_{\mathcal{E}_2},\ldots} &\sum_{ijk} | m_{ijk}  + \langle \langle M_0 \vert 2\mathcal{R}_k \mathcal{R}_j \mathcal{R}_i -  \tilde{\mathcal{R}}_{\mathcal{E}_k}\mathcal{R}_k \mathcal{R}_j \mathcal{R}_i\\ 
& -    \mathcal{R}_k \tilde{\mathcal{R}}_{\mathcal{E}_j}\mathcal{R}_j \mathcal{R}_i - \mathcal{R}_k \mathcal{R}_j \tilde{\mathcal{R}}_{\mathcal{E}_i}\mathcal{R}_i \vert \rho_0 \rangle \rangle|^2.
\end{split}
\end{equation}
subject to the constraints that the error maps $\tilde{\mathcal{R}}_{\mathcal{E}_j}$ are all physical.  

Minimizing this linearized least-square fit under the physicality constraint is another example of a semidefinite program.  In fact, it is isomorphic to the original QPT problem on a larger space where $\mathcal{R}_\Lambda \rightarrow R_{\mathcal{E}_1} \oplus R_{\mathcal{E}_2}\oplus \ldots R_{\mathcal{E}_N}$.  There is however one major difference which is that the optimal fit is no longer unique, which implies that the matrix corresponding to $SS^T$ in this space is not full-rank.  As an example take any unitary transformation $G^{(\rm ideal)} = \{  \mathcal{R}_1 , \mathcal{R}_2,\ldots   \mathcal{R}_N\} \rightarrow G^{(\rm err)} = \{  \mathcal{R}^T_{U}\mathcal{R}_1  \mathcal{R}_{U},  \mathcal{R}^T_{U}\mathcal{R}_2\mathcal{R}_{U},\ldots    \mathcal{R}^T_{U}\mathcal{R}_N\mathcal{R}_{U} \}$ where $\mathcal{R}_{U} \vert \rho_0 \rangle \rangle  \langle \langle M_0 \vert \mathcal{R}^T_{U} = \vert \rho_0 \rangle \rangle  \langle \langle M_0 \vert$. The measurement outcomes are invariant to such a transformation since
\begin{equation}
\begin{split}
m_{ijk} =& {\rm Tr} \Big(\vert \rho_0 \rangle \rangle  \langle \langle M_0 \vert \mathcal{R}_i\mathcal{R}_j\mathcal{R}_k \Big)\\
\rightarrow&{\rm Tr} \Big(\vert \rho_0 \rangle \rangle  \langle \langle M_0 \vert \mathcal{R}^T_{U} \mathcal{R}_i\mathcal{R}_U \mathcal{R}^T_{U}\mathcal{R}_j \mathcal{R}_U\mathcal{R}^T_{U}\mathcal{R}_k \mathcal{R}_U\Big)\\
=&{\rm Tr}\Big(\mathcal{R}_U\vert \rho_0 \rangle \rangle  \langle \langle M_0 \vert \mathcal{R}^T_{U} \mathcal{R}_i\mathcal{R}_j \mathcal{R}\mathcal{R}_k \Big) = m_{ijk}
\end{split}
\end{equation}  
When we prepare and measure in the computational basis this is equivalent to a frame invariance with respect to a diagonal unitary transformation.  This is not analogous to moving to the interaction picture since the resulting frame is stationary, and does not generate an inertial correction to the Hamiltonian.  It is more akin to an energy rescaling.  Such a frame is  physically irrelevant and in the following simulations we calculate distance between gate-sets by optimizing to find the most generous frame.

In Fig.~\ref{fig:overkill} we compare the results of standard QPT and our self-consistent approach.  The error model is that of individual random unitary errors, the uncorrectable error from the previous section, and the gate library is the six element gate-set that maps $\vert 0 \rangle$ to the cardinal directions on the Bloch sphere.  We plot the errors from both the fidelity and the diamond norm of the estimates versus the error of the physical gate-set with respect to the ideal target.  For each we show the average result for the six gates.  Additionally, we also plot the ratio of the error in the estimate to the ideal for both measures.  

In both cases there appears to be an exponential improvement in accuracy over standard QPT with respect to decreasing gate error.  This is fundamentally different from the results of the previous section where the fidelity error changed substantially but the diamond norm remained qualitatively the same, and so this method is not an effective depolarization of the error.

To clarify the differences between the two approaches we plot the least square estimator Eq.~\eqref{eq:gen_likely} for the two estimates in  Fig.~\ref{fig:likely}.  Again we look at the ratio of the value of the least square fit for the estimated value divided by the fit of the ideal gates to the data.  The estimate from QPT is flat with respect to lowering gate error which corresponds to the fact that QPT is minimizing the wrong quantity.  Our estimator does not properly minimize this quantity either, due to the linear approximation, but the scaling is dramatically more favourable.  

There are some obvious shortcomings of this method.  We have increased the number of experiments and the complexity of the reconstruction algorithm.  These increases are both polynomial in the dimension of the system, but since the dimension is already exponential in the number of qubits the resulting overhead is non-trivial.  In principle, we can apply these techniques to two or three qubit systems. However, the reconstructions in  Fig.~\ref{fig:overkill} are already taxing the default numerical precision of our semidefinite optimizer.  If we extended the plot in Fig.~\ref{fig:likely} to the left we would very quickly see the self-consistent method flatten from the numerical tolerance.  Also, the maximization of diagonal frames is computationally expensive, generally more so than the SDP solver itself.

\section{Experiment}
\label{sec:Experiment}

\begin{table*}%
\begin{ruledtabular}
\begin{tabular}{|c|c|c|c|c|c|c|}
Sample & Gate& $1-F_g(\Lambda_{\rm QPT},\Lambda_{\rm ideal})$ & $1-F_g(\Lambda_{\rm SC},\Lambda_{\rm ideal})$& $1-F_g$ (RB)  & $\|\Lambda_{\rm QPT}-\Lambda_{\rm ideal} \|_{\Diamond}$ & $\|\Lambda_{\rm SC}-\Lambda_{\rm ideal} \|_{\Diamond}$   \\
\hline\hline
A & $I$ & 0.0058 & 0.0051 &-& 0.037 & 0.029\\
A & $X_{\pi}$ & 0.013 & 0.0098 &-& 0.075 & 0.042\\
A & $X_{\pi/2}$ & 0.0077 & 0.0047 &-& 0.049 & 0.018\\
A & $Y_{\pi/2}$ & 0.0096 & 0.0053 &-& 0.069 & 0.068\\
\hline
A & $\langle U \rangle $ & 0.0090 & 0.0062 &0.0029 & 0.057 & 0.039\\
\hline\hline
B & $I$ & 0.11 & 0.037 &-& 0.43 & 0.13\\
B & $X_{\pi}$ & 0.039 & 0.0057 &-& 0.31 & 0.041\\
B & $X_{\pi/2}$ & 0.040 & 0.014 &-& 0.37 & 0.090\\
B & $Y_{\pi/2}$ & 0.032 & 0.021 &-& 0.32 & 0.090\\
\hline
B & $\langle U \rangle$ & 0.057 & 0.020 & 0.0016 & 0.36 & 0.087\\
\end{tabular}%
\end{ruledtabular}
\caption{The gate error from QPT and self-consistent tomography for A) a SJT capacitively coupled to a coplanar waveguide  and B) a SJT suspended in a copper cavity.  We report the fidelity error and diamond norm for both approaches (QPT and self-consistent tomography), and for each gate in the set.  The last row in each sample, labeled $\langle U \rangle$, is the average of the previous four as well as the average error obtained from randomized benchmarking.  }\label{fidtab}
\end{table*}

\begin{figure*}
\centering
\includegraphics[width=0.9\textwidth]{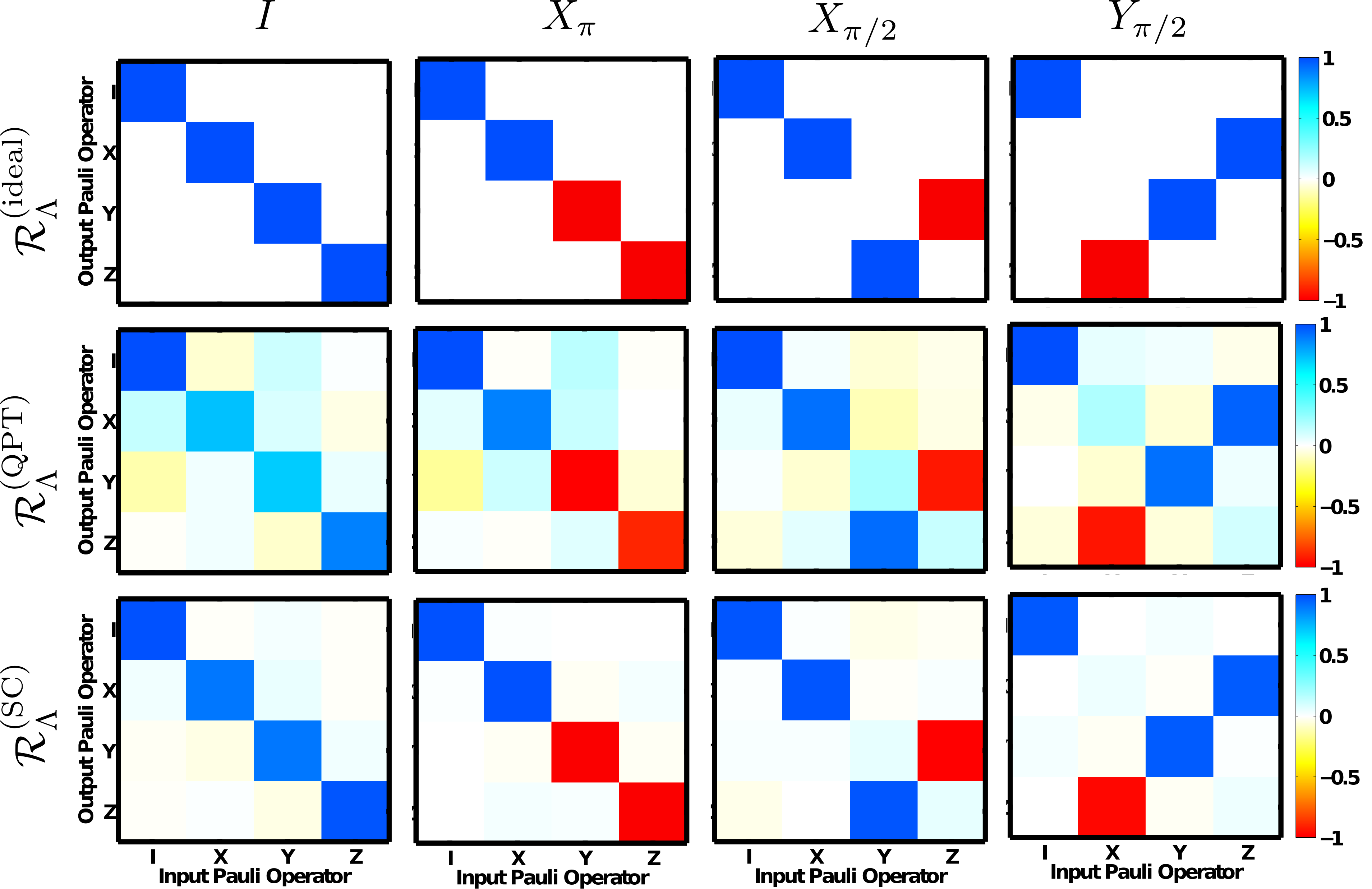}
\caption{(color online)  The $\mathcal{R}$-matrices for the ideal gates and the reconstruction from QPT and self-consistent tomography for each of the four measured gates on sample B.}
\label{fig:raulis}
\end{figure*}

We apply our self consistent tomography method to data generated from two independent experiments with single junction transmon (SJT) qubits.  Sample A is a SJT capacitively coupled to a coplanar waveguide resonator (used also in reference \cite{gambetta2012} as sample b), while sample B is mounted in a cavity machined from bulk copper.  In both cases the qubits are driven by resonant microwaves that are shaped in order to perform optimal qubit gates.  For these experiments we use the gate-set $\{ I, X_{\pi}, X_{\pi/2}, Y_{\pi/2}\}$ where the notation $R_{\theta}$ denotes a rotation about the axis $R$ of angle $\theta$.  See references \cite{Chow2012,rigetti2012} for the design and construction of these systems.

In sample A(B) the qubit frequency was $\omega_q /2\pi = 4.7610 (4.4738 ) {\rm GHz} $ and was coupled to a resonator of frequency $\omega_r/2\pi = 7.4269 (12.118 ){\rm GHz}$.  In sample A there was a second qubit coupled to the resonator with frequency $\omega_q /2\pi = 5.3401$ that played no role in the following experiment since the two qubits were only weakly coupled.  The coherence times of the two samples were $T_1 = 8.1 (44.0 ) \mu {\rm s}$ and $T_2^* = 16.2(7.2)\mu {\rm s} $.  The samples are radiation shielded and cooled to $15$mK in a dilution refrigerator.  
 
In Table \ref{fidtab} we show the results of both estimation methods as compared to the desired gate.  In both cases we find that QPT overestimates the error in both the diamond norm and the gate fidelity as compared to the self-consistent method.  For sample A the error on the QPT was roughly $1.5$ times the self-consistent error while for sample B the it was roughly 3 times as large.  We show plots of the Pauli transfer matrices for both estimates and the ideal maps in Fig.~\ref{fig:raulis} for sample B.  

Furthermore, since the gates in our library were generators of the Clifford group we were able to perform randomized benchmarking on the two samples using the methods in \cite{Magesan2011,Magesan2012}.  The fidelity error from randomized benchmarking was even smaller than the error from the self-consistent reconstruction, dramatically so in the case of sample B.  We conjecture that this is due to errors in the measurement operator $M_0$, since the calibration of the measurement voltage to the $\ket{0}$ and $\ket{1}$ states has been observed to shift over the course of long experimental runs.  Randomized benchmarking is insensitive to this type of error while the methods presented in this manuscript are not.  Accounting for errors in the initial state $\rho_0$ and the fixed observable $M_0$ are both important future avenues to extend this method, as well as that of dealing with slowly time-varying errors.

\section{Conclusion}
The methods of tomography are very powerful when using a well-characterized system to probe a black-box operation, but this is rarely the case when verifying quantum computing hardware.  Instead, the situation is one in which all components are faulty, and in that scenario we have shown that process tomography can fail dramatically.  Instead of separating the preparation and measurement phases of tomography we instead must view the experiment as a sequence of faulty gates, and while we have not discussed errors on the fiducial state and fixed measurement operator, those too should probably be brought into question in the future.  Without incorporating these errors into our model it is impossible to obtain a more accurate reconstruction through statistical analysis alone.

In this manuscript we have developed a new protocol that self-consistently reconstructs a library of quantum gates by modifying the likelihood function of quantum process tomography to incorporate our uncertainty of the state preparation and measurement gates.  In simulation, this self-consistent method outperforms standard QPT in terms of the accuracy of estimation since QPT consistently underestimates gate fidelities in the presence of SPAM errors.  This method requires the same number of experiments as QPT and only adds polynomial overhead to the amount of classical post-processing.  In fact, in any case where tomography has been performed on the entire set of gates used for state preparation and measurement no further experiments are needed to apply our protocol.  Colloquially, we have taken to calling this the `overkill' method of tomography, since the number of the experiments borders on the ridiculous for all but one or two qubit systems, but this will be the case for any protocol that provides full information about a library of maps.  Self-consistently reconstructing a library of gates, as opposed to performing tomography on the individual members, yields more trustworthy and generally higher fidelity estimates in both experiment and simulation and we expect the difference between standard QPT and the self-consistent method presented here to grow more pronounced as we reach lower and lower gate errors.    

\begin{acknowledgments}
We acknowledge contributions from J. Chow and Thomas Ohki and discussions with R. Blume-Kohout and Graeme Smith. We acknowledge support from IARPA under contract W911NF-10-1-0324. All statements of fact, opinion or conclusions contained herein are those of the authors and should not be construed as representing the official views or policies of the U.S. Government.
\end{acknowledgments}

\bibliography{SCtomo}

\begin{thebibliography}{31}%
\makeatletter
\providecommand \@ifxundefined [1]{%
 \@ifx{#1\undefined}
}%
\providecommand \@ifnum [1]{%
 \ifnum #1\expandafter \@firstoftwo
 \else \expandafter \@secondoftwo
 \fi
}%
\providecommand \@ifx [1]{%
 \ifx #1\expandafter \@firstoftwo
 \else \expandafter \@secondoftwo
 \fi
}%
\providecommand \natexlab [1]{#1}%
\providecommand \enquote  [1]{``#1''}%
\providecommand \bibnamefont  [1]{#1}%
\providecommand \bibfnamefont [1]{#1}%
\providecommand \citenamefont [1]{#1}%
\providecommand \href@noop [0]{\@secondoftwo}%
\providecommand \href [0]{\begingroup \@sanitize@url \@href}%
\providecommand \@href[1]{\@@startlink{#1}\@@href}%
\providecommand \@@href[1]{\endgroup#1\@@endlink}%
\providecommand \@sanitize@url [0]{\catcode `\\12\catcode `\$12\catcode
  `\&12\catcode `\#12\catcode `\^12\catcode `\_12\catcode `\%12\relax}%
\providecommand \@@startlink[1]{}%
\providecommand \@@endlink[0]{}%
\providecommand \url  [0]{\begingroup\@sanitize@url \@url }%
\providecommand \@url [1]{\endgroup\@href {#1}{\urlprefix }}%
\providecommand \urlprefix  [0]{URL }%
\providecommand \Eprint [0]{\href }%
\providecommand \doibase [0]{http://dx.doi.org/}%
\providecommand \selectlanguage [0]{\@gobble}%
\providecommand \bibinfo  [0]{\@secondoftwo}%
\providecommand \bibfield  [0]{\@secondoftwo}%
\providecommand \translation [1]{[#1]}%
\providecommand \BibitemOpen [0]{}%
\providecommand \bibitemStop [0]{}%
\providecommand \bibitemNoStop [0]{.\EOS\space}%
\providecommand \EOS [0]{\spacefactor3000\relax}%
\providecommand \BibitemShut  [1]{\csname bibitem#1\endcsname}%
\let\auto@bib@innerbib\@empty
\bibitem [{\citenamefont {Chow}\ \emph {et~al.}(2012)\citenamefont {Chow},
  \citenamefont {Gambetta}, \citenamefont {C\'orcoles}, \citenamefont {Merkel},
  \citenamefont {Smolin}, \citenamefont {Rigetti}, \citenamefont {Poletto},
  \citenamefont {Keefe}, \citenamefont {Rothwell}, \citenamefont {Rozen},
  \citenamefont {Ketchen},\ and\ \citenamefont {Steffen}}]{Chow2012}%
  \BibitemOpen
  \bibfield  {author} {\bibinfo {author} {\bibfnamefont {J.~M.}\ \bibnamefont
  {Chow}}, \bibinfo {author} {\bibfnamefont {J.~M.}\ \bibnamefont {Gambetta}},
  \bibinfo {author} {\bibfnamefont {A.~D.}\ \bibnamefont {C\'orcoles}},
  \bibinfo {author} {\bibfnamefont {S.~T.}\ \bibnamefont {Merkel}}, \bibinfo
  {author} {\bibfnamefont {J.~A.}\ \bibnamefont {Smolin}}, \bibinfo {author}
  {\bibfnamefont {C.}~\bibnamefont {Rigetti}}, \bibinfo {author} {\bibfnamefont
  {S.}~\bibnamefont {Poletto}}, \bibinfo {author} {\bibfnamefont {G.~A.}\
  \bibnamefont {Keefe}}, \bibinfo {author} {\bibfnamefont {M.~B.}\ \bibnamefont
  {Rothwell}}, \bibinfo {author} {\bibfnamefont {J.~R.}\ \bibnamefont {Rozen}},
  \bibinfo {author} {\bibfnamefont {M.~B.}\ \bibnamefont {Ketchen}}, \ and\
  \bibinfo {author} {\bibfnamefont {M.}~\bibnamefont {Steffen}},\ }\href
  {\doibase 10.1103/PhysRevLett.109.060501} {\bibfield  {journal} {\bibinfo
  {journal} {Phys. Rev. Lett.}\ }\textbf {\bibinfo {volume} {109}},\ \bibinfo
  {pages} {060501} (\bibinfo {year} {2012})}\BibitemShut {NoStop}%
\bibitem [{\citenamefont {DiCarlo}\ \emph {et~al.}(2009)\citenamefont
  {DiCarlo}, \citenamefont {Chow}, \citenamefont {Gambetta}, \citenamefont
  {Bishop}, \citenamefont {Johnson}, \citenamefont {Schuster}, \citenamefont
  {Majer}, \citenamefont {Blais}, \citenamefont {Frunzio}, \citenamefont
  {Girvin},\ and\ \citenamefont {Schoelkopf}}]{dicarlo_2009}%
  \BibitemOpen
  \bibfield  {author} {\bibinfo {author} {\bibfnamefont {L.}~\bibnamefont
  {DiCarlo}}, \bibinfo {author} {\bibfnamefont {J.~M.}\ \bibnamefont {Chow}},
  \bibinfo {author} {\bibfnamefont {J.~M.}\ \bibnamefont {Gambetta}}, \bibinfo
  {author} {\bibfnamefont {L.~S.}\ \bibnamefont {Bishop}}, \bibinfo {author}
  {\bibfnamefont {B.~R.}\ \bibnamefont {Johnson}}, \bibinfo {author}
  {\bibfnamefont {D.~I.}\ \bibnamefont {Schuster}}, \bibinfo {author}
  {\bibfnamefont {J.}~\bibnamefont {Majer}}, \bibinfo {author} {\bibfnamefont
  {A.}~\bibnamefont {Blais}}, \bibinfo {author} {\bibfnamefont
  {L.}~\bibnamefont {Frunzio}}, \bibinfo {author} {\bibfnamefont {S.~M.}\
  \bibnamefont {Girvin}}, \ and\ \bibinfo {author} {\bibfnamefont {R.~J.}\
  \bibnamefont {Schoelkopf}},\ }\href {\doibase 10.1038/nature08121} {\bibfield
   {journal} {\bibinfo  {journal} {Nature}\ }\textbf {\bibinfo {volume}
  {460}},\ \bibinfo {pages} {240} (\bibinfo {year} {2009})}\BibitemShut
  {NoStop}%
\bibitem [{\citenamefont {Bialczak}\ \emph {et~al.}(2010)\citenamefont
  {Bialczak}, \citenamefont {Ansmann}, \citenamefont {Hofheinz}, \citenamefont
  {Lucero}, \citenamefont {Neeley}, \citenamefont {O'Connell}, \citenamefont
  {Sank}, \citenamefont {Wang}, \citenamefont {Wenner}, \citenamefont
  {Steffen}, \citenamefont {Cleland},\ and\ \citenamefont
  {Martinis}}]{Bialczak2010}%
  \BibitemOpen
  \bibfield  {author} {\bibinfo {author} {\bibfnamefont {R.~C.}\ \bibnamefont
  {Bialczak}}, \bibinfo {author} {\bibfnamefont {M.}~\bibnamefont {Ansmann}},
  \bibinfo {author} {\bibfnamefont {M.}~\bibnamefont {Hofheinz}}, \bibinfo
  {author} {\bibfnamefont {E.}~\bibnamefont {Lucero}}, \bibinfo {author}
  {\bibfnamefont {M.}~\bibnamefont {Neeley}}, \bibinfo {author} {\bibfnamefont
  {A.~D.}\ \bibnamefont {O'Connell}}, \bibinfo {author} {\bibfnamefont
  {D.}~\bibnamefont {Sank}}, \bibinfo {author} {\bibfnamefont {H.}~\bibnamefont
  {Wang}}, \bibinfo {author} {\bibfnamefont {J.}~\bibnamefont {Wenner}},
  \bibinfo {author} {\bibfnamefont {M.}~\bibnamefont {Steffen}}, \bibinfo
  {author} {\bibfnamefont {A.~N.}\ \bibnamefont {Cleland}}, \ and\ \bibinfo
  {author} {\bibfnamefont {J.~M.}\ \bibnamefont {Martinis}},\ }\href
  {http://dx.doi.org/10.1038/nphys1639} {\bibfield  {journal} {\bibinfo
  {journal} {Nat Phys}\ }\textbf {\bibinfo {volume} {6}},\ \bibinfo {pages}
  {409} (\bibinfo {year} {2010})}\BibitemShut {NoStop}%
\bibitem [{\citenamefont {Haffner}\ \emph {et~al.}(2005)\citenamefont
  {Haffner}, \citenamefont {Hansel}, \citenamefont {Roos}, \citenamefont
  {Benhelm}, \citenamefont {Chek-al kar}, \citenamefont {Chwalla},
  \citenamefont {Korber}, \citenamefont {Rapol}, \citenamefont {Riebe},
  \citenamefont {Schmidt}, \citenamefont {Becher}, \citenamefont {Guhne},
  \citenamefont {Dur},\ and\ \citenamefont
  {Blatt}}]{Haffner_multiparticle_ions_2005}%
  \BibitemOpen
  \bibfield  {author} {\bibinfo {author} {\bibfnamefont {H.}~\bibnamefont
  {Haffner}}, \bibinfo {author} {\bibfnamefont {W.}~\bibnamefont {Hansel}},
  \bibinfo {author} {\bibfnamefont {C.~F.}\ \bibnamefont {Roos}}, \bibinfo
  {author} {\bibfnamefont {J.}~\bibnamefont {Benhelm}}, \bibinfo {author}
  {\bibfnamefont {D.}~\bibnamefont {Chek-al kar}}, \bibinfo {author}
  {\bibfnamefont {M.}~\bibnamefont {Chwalla}}, \bibinfo {author} {\bibfnamefont
  {T.}~\bibnamefont {Korber}}, \bibinfo {author} {\bibfnamefont {U.~D.}\
  \bibnamefont {Rapol}}, \bibinfo {author} {\bibfnamefont {M.}~\bibnamefont
  {Riebe}}, \bibinfo {author} {\bibfnamefont {P.~O.}\ \bibnamefont {Schmidt}},
  \bibinfo {author} {\bibfnamefont {C.}~\bibnamefont {Becher}}, \bibinfo
  {author} {\bibfnamefont {O.}~\bibnamefont {Guhne}}, \bibinfo {author}
  {\bibfnamefont {W.}~\bibnamefont {Dur}}, \ and\ \bibinfo {author}
  {\bibfnamefont {R.}~\bibnamefont {Blatt}},\ }\href {\doibase
  10.1038/nature04279} {\bibfield  {journal} {\bibinfo  {journal} {Nature}\
  }\textbf {\bibinfo {volume} {438}},\ \bibinfo {pages} {643} (\bibinfo {year}
  {2005})}\BibitemShut {NoStop}%
\bibitem [{\citenamefont {Gaebler}\ \emph {et~al.}(2012)\citenamefont
  {Gaebler}, \citenamefont {Meier}, \citenamefont {Tan}, \citenamefont
  {Bowler}, \citenamefont {Lin}, \citenamefont {Hanneke}, \citenamefont {Jost},
  \citenamefont {Home}, \citenamefont {Knill}, \citenamefont {Leibfried},\ and\
  \citenamefont {Wineland}}]{gaebler2012}%
  \BibitemOpen
  \bibfield  {author} {\bibinfo {author} {\bibfnamefont {J.~P.}\ \bibnamefont
  {Gaebler}}, \bibinfo {author} {\bibfnamefont {A.~M.}\ \bibnamefont {Meier}},
  \bibinfo {author} {\bibfnamefont {T.~R.}\ \bibnamefont {Tan}}, \bibinfo
  {author} {\bibfnamefont {R.}~\bibnamefont {Bowler}}, \bibinfo {author}
  {\bibfnamefont {Y.}~\bibnamefont {Lin}}, \bibinfo {author} {\bibfnamefont
  {D.}~\bibnamefont {Hanneke}}, \bibinfo {author} {\bibfnamefont {J.~D.}\
  \bibnamefont {Jost}}, \bibinfo {author} {\bibfnamefont {J.~P.}\ \bibnamefont
  {Home}}, \bibinfo {author} {\bibfnamefont {E.}~\bibnamefont {Knill}},
  \bibinfo {author} {\bibfnamefont {D.}~\bibnamefont {Leibfried}}, \ and\
  \bibinfo {author} {\bibfnamefont {D.~J.}\ \bibnamefont {Wineland}},\ }\href
  {\doibase 10.1103/PhysRevLett.108.260503} {\bibfield  {journal} {\bibinfo
  {journal} {Phys. Rev. Lett.}\ }\textbf {\bibinfo {volume} {108}},\ \bibinfo
  {pages} {260503} (\bibinfo {year} {2012})}\BibitemShut {NoStop}%
\bibitem [{\citenamefont {Nielsen}(2002)}]{nielsen_gatefid_2002}%
  \BibitemOpen
  \bibfield  {author} {\bibinfo {author} {\bibfnamefont {M.~A.}\ \bibnamefont
  {Nielsen}},\ }\href {\doibase 10.1016/S0375-9601(02)01272-0} {\bibfield
  {journal} {\bibinfo  {journal} {Phys. Lett. A}\ }\textbf {\bibinfo {volume}
  {303}},\ \bibinfo {pages} {249} (\bibinfo {year} {2002})}\BibitemShut
  {NoStop}%
\bibitem [{\citenamefont {Chuang}\ and\ \citenamefont
  {Nielsen}(1997)}]{chuang_blackbox_1997}%
  \BibitemOpen
  \bibfield  {author} {\bibinfo {author} {\bibfnamefont {I.~L.}\ \bibnamefont
  {Chuang}}\ and\ \bibinfo {author} {\bibfnamefont {M.}~\bibnamefont
  {Nielsen}},\ }\href {\doibase 10.1080/095003497152609} {\bibfield  {journal}
  {\bibinfo  {journal} {Journal of Modern Optics}\ }\textbf {\bibinfo {volume}
  {44}},\ \bibinfo {pages} {2455} (\bibinfo {year} {1997})}\BibitemShut
  {NoStop}%
\bibitem [{\citenamefont {Gambetta}\ \emph {et~al.}(2012)\citenamefont
  {Gambetta}, \citenamefont {Corcoles}, \citenamefont {Merkel}, \citenamefont
  {Johnson}, \citenamefont {Smolin}, \citenamefont {Chow}, \citenamefont
  {Ryan}, \citenamefont {Rigetti}, \citenamefont {Poletto}, \citenamefont
  {Ohki}, \citenamefont {Ketchen},\ and\ \citenamefont
  {Steffen}}]{gambetta2012}%
  \BibitemOpen
  \bibfield  {author} {\bibinfo {author} {\bibfnamefont {J.~M.}\ \bibnamefont
  {Gambetta}}, \bibinfo {author} {\bibfnamefont {A.~D.}\ \bibnamefont
  {Corcoles}}, \bibinfo {author} {\bibfnamefont {S.~T.}\ \bibnamefont
  {Merkel}}, \bibinfo {author} {\bibfnamefont {B.~R.}\ \bibnamefont {Johnson}},
  \bibinfo {author} {\bibfnamefont {J.~A.}\ \bibnamefont {Smolin}}, \bibinfo
  {author} {\bibfnamefont {J.~M.}\ \bibnamefont {Chow}}, \bibinfo {author}
  {\bibfnamefont {C.~A.}\ \bibnamefont {Ryan}}, \bibinfo {author}
  {\bibfnamefont {C.}~\bibnamefont {Rigetti}}, \bibinfo {author} {\bibfnamefont
  {S.}~\bibnamefont {Poletto}}, \bibinfo {author} {\bibfnamefont {T.~A.}\
  \bibnamefont {Ohki}}, \bibinfo {author} {\bibfnamefont {M.~B.}\ \bibnamefont
  {Ketchen}}, \ and\ \bibinfo {author} {\bibfnamefont {M.}~\bibnamefont
  {Steffen}},\ }\href@noop {} {\bibfield  {journal} {\bibinfo  {journal}
  {arXiv:1204.6308}\ } (\bibinfo {year} {2012})}\BibitemShut {NoStop}%
\bibitem [{\citenamefont {James}\ \emph {et~al.}(2001)\citenamefont {James},
  \citenamefont {Kwiat}, \citenamefont {Munro},\ and\ \citenamefont
  {White}}]{James_qubitmeas_2001}%
  \BibitemOpen
  \bibfield  {author} {\bibinfo {author} {\bibfnamefont {D.~F.~V.}\
  \bibnamefont {James}}, \bibinfo {author} {\bibfnamefont {P.~G.}\ \bibnamefont
  {Kwiat}}, \bibinfo {author} {\bibfnamefont {W.~J.}\ \bibnamefont {Munro}}, \
  and\ \bibinfo {author} {\bibfnamefont {A.~G.}\ \bibnamefont {White}},\ }\href
  {\doibase 10.1103/PhysRevA.64.052312} {\bibfield  {journal} {\bibinfo
  {journal} {Phys. Rev. A}\ }\textbf {\bibinfo {volume} {64}},\ \bibinfo
  {pages} {052312} (\bibinfo {year} {2001})}\BibitemShut {NoStop}%
\bibitem [{\citenamefont {Hradil}\ \emph {et~al.}(2004)\citenamefont {Hradil},
  \citenamefont {Rehacek}, \citenamefont {Fiurasek},\ and\ \citenamefont
  {Jezek}}]{Hradil2004}%
  \BibitemOpen
  \bibfield  {author} {\bibinfo {author} {\bibfnamefont {Z.}~\bibnamefont
  {Hradil}}, \bibinfo {author} {\bibfnamefont {J.}~\bibnamefont {Rehacek}},
  \bibinfo {author} {\bibfnamefont {J.}~\bibnamefont {Fiurasek}}, \ and\
  \bibinfo {author} {\bibfnamefont {M.}~\bibnamefont {Jezek}},\ }\href@noop {}
  {\bibfield  {journal} {\bibinfo  {journal} {Lect. Notes Phys.}\ }\textbf
  {\bibinfo {volume} {649}},\ \bibinfo {pages} {59} (\bibinfo {year}
  {2004})}\BibitemShut {NoStop}%
\bibitem [{\citenamefont {Lvovsky}(2004)}]{lvosky04}%
  \BibitemOpen
  \bibfield  {author} {\bibinfo {author} {\bibfnamefont {A.~I.}\ \bibnamefont
  {Lvovsky}},\ }\href {http://stacks.iop.org/1464-4266/6/i=6/a=014} {\bibfield
  {journal} {\bibinfo  {journal} {Journal of Optics B: Quantum and
  Semiclassical Optics}\ }\textbf {\bibinfo {volume} {6}},\ \bibinfo {pages}
  {S556} (\bibinfo {year} {2004})}\BibitemShut {NoStop}%
\bibitem [{\citenamefont {de~Burgh}\ \emph {et~al.}(2008)\citenamefont
  {de~Burgh}, \citenamefont {Langford}, \citenamefont {Doherty},\ and\
  \citenamefont {Gilchrist}}]{deBrugh2008}%
  \BibitemOpen
  \bibfield  {author} {\bibinfo {author} {\bibfnamefont {M.~D.}\ \bibnamefont
  {de~Burgh}}, \bibinfo {author} {\bibfnamefont {N.~K.}\ \bibnamefont
  {Langford}}, \bibinfo {author} {\bibfnamefont {A.~C.}\ \bibnamefont
  {Doherty}}, \ and\ \bibinfo {author} {\bibfnamefont {A.}~\bibnamefont
  {Gilchrist}},\ }\href {\doibase 10.1103/PhysRevA.78.052122} {\bibfield
  {journal} {\bibinfo  {journal} {Phys. Rev. A}\ }\textbf {\bibinfo {volume}
  {78}},\ \bibinfo {pages} {052122} (\bibinfo {year} {2008})}\BibitemShut
  {NoStop}%
\bibitem [{\citenamefont {Blume-Kohout}(2010{\natexlab{a}})}]{rbk2010}%
  \BibitemOpen
  \bibfield  {author} {\bibinfo {author} {\bibfnamefont {R.}~\bibnamefont
  {Blume-Kohout}},\ }\href {\doibase 10.1103/PhysRevLett.105.200504} {\bibfield
   {journal} {\bibinfo  {journal} {Phys. Rev. Lett.}\ }\textbf {\bibinfo
  {volume} {105}},\ \bibinfo {pages} {200504} (\bibinfo {year}
  {2010}{\natexlab{a}})}\BibitemShut {NoStop}%
\bibitem [{\citenamefont {Blume-Kohout}(2010{\natexlab{b}})}]{blume2012}%
  \BibitemOpen
  \bibfield  {author} {\bibinfo {author} {\bibfnamefont {R.}~\bibnamefont
  {Blume-Kohout}},\ }\href {http://stacks.iop.org/1367-2630/12/i=4/a=043034}
  {\bibfield  {journal} {\bibinfo  {journal} {New Journal of Physics}\ }\textbf
  {\bibinfo {volume} {12}},\ \bibinfo {pages} {043034} (\bibinfo {year}
  {2010}{\natexlab{b}})}\BibitemShut {NoStop}%
\bibitem [{\citenamefont {Smolin}\ \emph {et~al.}(2012)\citenamefont {Smolin},
  \citenamefont {Gambetta},\ and\ \citenamefont {Smith}}]{smollin2012}%
  \BibitemOpen
  \bibfield  {author} {\bibinfo {author} {\bibfnamefont {J.~A.}\ \bibnamefont
  {Smolin}}, \bibinfo {author} {\bibfnamefont {J.~M.}\ \bibnamefont
  {Gambetta}}, \ and\ \bibinfo {author} {\bibfnamefont {G.}~\bibnamefont
  {Smith}},\ }\href {\doibase 10.1103/PhysRevLett.108.070502} {\bibfield
  {journal} {\bibinfo  {journal} {Phys. Rev. Lett.}\ }\textbf {\bibinfo
  {volume} {108}},\ \bibinfo {pages} {070502} (\bibinfo {year}
  {2012})}\BibitemShut {NoStop}%
\bibitem [{\citenamefont {Christandl}\ and\ \citenamefont
  {Renner}(2012)}]{Christandl2012}%
  \BibitemOpen
  \bibfield  {author} {\bibinfo {author} {\bibfnamefont {M.}~\bibnamefont
  {Christandl}}\ and\ \bibinfo {author} {\bibfnamefont {R.}~\bibnamefont
  {Renner}},\ }\href {\doibase 10.1103/PhysRevLett.109.120403} {\bibfield
  {journal} {\bibinfo  {journal} {Phys. Rev. Lett.}\ }\textbf {\bibinfo
  {volume} {109}},\ \bibinfo {pages} {120403} (\bibinfo {year}
  {2012})}\BibitemShut {NoStop}%
\bibitem [{\citenamefont {Knill}\ \emph {et~al.}(2008)\citenamefont {Knill},
  \citenamefont {Leibfried}, \citenamefont {Reichle}, \citenamefont {Britton},
  \citenamefont {Blakestad}, \citenamefont {Jost}, \citenamefont {Langer},
  \citenamefont {Ozeri}, \citenamefont {Seidelin},\ and\ \citenamefont
  {Wineland}}]{knill_randomized_2008}%
  \BibitemOpen
  \bibfield  {author} {\bibinfo {author} {\bibfnamefont {E.}~\bibnamefont
  {Knill}}, \bibinfo {author} {\bibfnamefont {D.}~\bibnamefont {Leibfried}},
  \bibinfo {author} {\bibfnamefont {R.}~\bibnamefont {Reichle}}, \bibinfo
  {author} {\bibfnamefont {J.}~\bibnamefont {Britton}}, \bibinfo {author}
  {\bibfnamefont {R.~B.}\ \bibnamefont {Blakestad}}, \bibinfo {author}
  {\bibfnamefont {J.~D.}\ \bibnamefont {Jost}}, \bibinfo {author}
  {\bibfnamefont {C.}~\bibnamefont {Langer}}, \bibinfo {author} {\bibfnamefont
  {R.}~\bibnamefont {Ozeri}}, \bibinfo {author} {\bibfnamefont
  {S.}~\bibnamefont {Seidelin}}, \ and\ \bibinfo {author} {\bibfnamefont
  {D.~J.}\ \bibnamefont {Wineland}},\ }\href {\doibase
  10.1103/PhysRevA.77.012307} {\bibfield  {journal} {\bibinfo  {journal} {Phys.
  Rev. A}\ }\textbf {\bibinfo {volume} {77}},\ \bibinfo {pages} {012307}
  (\bibinfo {year} {2008})}\BibitemShut {NoStop}%
\bibitem [{\citenamefont {Magesan}\ \emph {et~al.}(2011)\citenamefont
  {Magesan}, \citenamefont {Gambetta},\ and\ \citenamefont
  {Emerson}}]{Magesan2011}%
  \BibitemOpen
  \bibfield  {author} {\bibinfo {author} {\bibfnamefont {E.}~\bibnamefont
  {Magesan}}, \bibinfo {author} {\bibfnamefont {J.~M.}\ \bibnamefont
  {Gambetta}}, \ and\ \bibinfo {author} {\bibfnamefont {J.}~\bibnamefont
  {Emerson}},\ }\href {http://link.aps.org/doi/10.1103/PhysRevLett.106.180504}
  {\bibfield  {journal} {\bibinfo  {journal} {Phys. Rev. Lett.}\ }\textbf
  {\bibinfo {volume} {106}},\ \bibinfo {pages} {180504} (\bibinfo {year}
  {2011})}\BibitemShut {NoStop}%
\bibitem [{\citenamefont {Magesan}\ \emph
  {et~al.}(2012{\natexlab{a}})\citenamefont {Magesan}, \citenamefont
  {Gambetta},\ and\ \citenamefont {Emerson}}]{Magesan2012}%
  \BibitemOpen
  \bibfield  {author} {\bibinfo {author} {\bibfnamefont {E.}~\bibnamefont
  {Magesan}}, \bibinfo {author} {\bibfnamefont {J.~M.}\ \bibnamefont
  {Gambetta}}, \ and\ \bibinfo {author} {\bibfnamefont {J.}~\bibnamefont
  {Emerson}},\ }\href {\doibase 10.1103/PhysRevA.85.042311} {\bibfield
  {journal} {\bibinfo  {journal} {Phys. Rev. A}\ }\textbf {\bibinfo {volume}
  {85}},\ \bibinfo {pages} {042311} (\bibinfo {year}
  {2012}{\natexlab{a}})}\BibitemShut {NoStop}%
\bibitem [{\citenamefont {Magesan}\ \emph
  {et~al.}(2012{\natexlab{b}})\citenamefont {Magesan}, \citenamefont
  {Gambetta}, \citenamefont {Johnson}, \citenamefont {Ryan}, \citenamefont
  {Chow}, \citenamefont {Merkel}, \citenamefont {da~Silva}, \citenamefont
  {Keefe}, \citenamefont {Rothwell}, \citenamefont {Ohki}, \citenamefont
  {Ketchen},\ and\ \citenamefont {Steffen}}]{Magesan2012b}%
  \BibitemOpen
  \bibfield  {author} {\bibinfo {author} {\bibfnamefont {E.}~\bibnamefont
  {Magesan}}, \bibinfo {author} {\bibfnamefont {J.~M.}\ \bibnamefont
  {Gambetta}}, \bibinfo {author} {\bibfnamefont {B.~R.}\ \bibnamefont
  {Johnson}}, \bibinfo {author} {\bibfnamefont {C.~A.}\ \bibnamefont {Ryan}},
  \bibinfo {author} {\bibfnamefont {J.~M.}\ \bibnamefont {Chow}}, \bibinfo
  {author} {\bibfnamefont {S.~T.}\ \bibnamefont {Merkel}}, \bibinfo {author}
  {\bibfnamefont {M.~P.}\ \bibnamefont {da~Silva}}, \bibinfo {author}
  {\bibfnamefont {G.~A.}\ \bibnamefont {Keefe}}, \bibinfo {author}
  {\bibfnamefont {M.~B.}\ \bibnamefont {Rothwell}}, \bibinfo {author}
  {\bibfnamefont {T.~A.}\ \bibnamefont {Ohki}}, \bibinfo {author}
  {\bibfnamefont {M.~B.}\ \bibnamefont {Ketchen}}, \ and\ \bibinfo {author}
  {\bibfnamefont {M.}~\bibnamefont {Steffen}},\ }\href {\doibase
  10.1103/PhysRevLett.109.080505} {\bibfield  {journal} {\bibinfo  {journal}
  {Phys. Rev. Lett.}\ }\textbf {\bibinfo {volume} {109}},\ \bibinfo {pages}
  {080505} (\bibinfo {year} {2012}{\natexlab{b}})}\BibitemShut {NoStop}%
\bibitem [{\citenamefont {Brańczyk}\ \emph {et~al.}(2012)\citenamefont
  {Brańczyk}, \citenamefont {Mahler}, \citenamefont {Rozema}, \citenamefont
  {Darabi}, \citenamefont {Steinberg},\ and\ \citenamefont
  {James}}]{james2012}%
  \BibitemOpen
  \bibfield  {author} {\bibinfo {author} {\bibfnamefont {A.~M.}\ \bibnamefont
  {Brańczyk}}, \bibinfo {author} {\bibfnamefont {D.~H.}\ \bibnamefont
  {Mahler}}, \bibinfo {author} {\bibfnamefont {L.~A.}\ \bibnamefont {Rozema}},
  \bibinfo {author} {\bibfnamefont {A.}~\bibnamefont {Darabi}}, \bibinfo
  {author} {\bibfnamefont {A.~M.}\ \bibnamefont {Steinberg}}, \ and\ \bibinfo
  {author} {\bibfnamefont {D.~F.~V.}\ \bibnamefont {James}},\ }\href
  {http://stacks.iop.org/1367-2630/14/i=8/a=085003} {\bibfield  {journal}
  {\bibinfo  {journal} {New Journal of Physics}\ }\textbf {\bibinfo {volume}
  {14}},\ \bibinfo {pages} {085003} (\bibinfo {year} {2012})}\BibitemShut
  {NoStop}%
\bibitem [{\citenamefont {Dobrovitski}\ \emph {et~al.}(2010)\citenamefont
  {Dobrovitski}, \citenamefont {de~Lange}, \citenamefont {Rist\`e},\ and\
  \citenamefont {Hanson}}]{Dobrovitski2010}%
  \BibitemOpen
  \bibfield  {author} {\bibinfo {author} {\bibfnamefont {V.~V.}\ \bibnamefont
  {Dobrovitski}}, \bibinfo {author} {\bibfnamefont {G.}~\bibnamefont
  {de~Lange}}, \bibinfo {author} {\bibfnamefont {D.}~\bibnamefont {Rist\`e}}, \
  and\ \bibinfo {author} {\bibfnamefont {R.}~\bibnamefont {Hanson}},\ }\href
  {\doibase 10.1103/PhysRevLett.105.077601} {\bibfield  {journal} {\bibinfo
  {journal} {Phys. Rev. Lett.}\ }\textbf {\bibinfo {volume} {105}},\ \bibinfo
  {pages} {077601} (\bibinfo {year} {2010})}\BibitemShut {NoStop}%
\bibitem [{\citenamefont {Choi}(1975)}]{Choi1975}%
  \BibitemOpen
  \bibfield  {author} {\bibinfo {author} {\bibfnamefont {M.-D.}\ \bibnamefont
  {Choi}},\ }\href
  {http://www.sciencedirect.com/science/article/pii/0024379575900750}
  {\bibfield  {journal} {\bibinfo  {journal} {Linear Algebra and its
  Applications}\ }\textbf {\bibinfo {volume} {10}},\ \bibinfo {pages} {285}
  (\bibinfo {year} {1975})}\BibitemShut {NoStop}%
\bibitem [{\citenamefont {Boyd}\ and\ \citenamefont {Vandenberghe}()}]{Boyd08}%
  \BibitemOpen
  \bibfield  {author} {\bibinfo {author} {\bibfnamefont {S.}~\bibnamefont
  {Boyd}}\ and\ \bibinfo {author} {\bibfnamefont {L.}~\bibnamefont
  {Vandenberghe}},\ }\href@noop {} {}\ (\bibinfo  {publisher} {Cambridge
  University Press})\BibitemShut {NoStop}%
\bibitem [{\citenamefont {Sturm}(1999)}]{sedumi}%
  \BibitemOpen
  \bibfield  {author} {\bibinfo {author} {\bibfnamefont {J.~F.}\ \bibnamefont
  {Sturm}},\ }\href@noop {} {\bibfield  {journal} {\bibinfo  {journal}
  {Optimization methods and Software}\ }\textbf {\bibinfo {volume} {11}},\
  \bibinfo {pages} {625} (\bibinfo {year} {1999})}\BibitemShut {NoStop}%
\bibitem [{\citenamefont {Silberfarb}\ \emph {et~al.}(2005)\citenamefont
  {Silberfarb}, \citenamefont {Jessen},\ and\ \citenamefont
  {Deutsch}}]{silberfarb05}%
  \BibitemOpen
  \bibfield  {author} {\bibinfo {author} {\bibfnamefont {A.}~\bibnamefont
  {Silberfarb}}, \bibinfo {author} {\bibfnamefont {P.~S.}\ \bibnamefont
  {Jessen}}, \ and\ \bibinfo {author} {\bibfnamefont {I.~H.}\ \bibnamefont
  {Deutsch}},\ }\href {\doibase 10.1103/PhysRevLett.95.030402} {\bibfield
  {journal} {\bibinfo  {journal} {Phys. Rev. Lett.}\ }\textbf {\bibinfo
  {volume} {95}},\ \bibinfo {pages} {030402} (\bibinfo {year}
  {2005})}\BibitemShut {NoStop}%
\bibitem [{\citenamefont {Riofrío}\ \emph {et~al.}(2011)\citenamefont
  {Riofrío}, \citenamefont {Jessen},\ and\ \citenamefont
  {Deutsch}}]{riofrio2011}%
  \BibitemOpen
  \bibfield  {author} {\bibinfo {author} {\bibfnamefont {C.~A.}\ \bibnamefont
  {Riofrío}}, \bibinfo {author} {\bibfnamefont {P.~S.}\ \bibnamefont
  {Jessen}}, \ and\ \bibinfo {author} {\bibfnamefont {I.~H.}\ \bibnamefont
  {Deutsch}},\ }\href {http://stacks.iop.org/0953-4075/44/i=15/a=154007}
  {\bibfield  {journal} {\bibinfo  {journal} {Journal of Physics B: Atomic,
  Molecular and Optical Physics}\ }\textbf {\bibinfo {volume} {44}},\ \bibinfo
  {pages} {154007} (\bibinfo {year} {2011})}\BibitemShut {NoStop}%
\bibitem [{\citenamefont {Kitaev}\ \emph {et~al.}(2002)\citenamefont {Kitaev},
  \citenamefont {Shen},\ and\ \citenamefont {Vyalyi}}]{kitaev02}%
  \BibitemOpen
  \bibfield  {author} {\bibinfo {author} {\bibfnamefont {A.~Y.}\ \bibnamefont
  {Kitaev}}, \bibinfo {author} {\bibfnamefont {A.}~\bibnamefont {Shen}}, \ and\
  \bibinfo {author} {\bibfnamefont {M.~N.}\ \bibnamefont {Vyalyi}},\
  }\href@noop {} {\emph {\bibinfo {title} {Classical and Quantum
  Computation}}}\ (\bibinfo  {publisher} {American Mathematical Society},\
  \bibinfo {year} {2002})\BibitemShut {NoStop}%
\bibitem [{\citenamefont {Watrous}()}]{watrous2009}%
  \BibitemOpen
  \bibfield  {author} {\bibinfo {author} {\bibfnamefont {J.}~\bibnamefont
  {Watrous}},\ }\href@noop {} {\bibfield  {journal} {\bibinfo  {journal}
  {Theory of Computing}\ }\textbf {\bibinfo {volume} {5}},\ \bibinfo {pages}
  {11}}\BibitemShut {NoStop}%
\bibitem [{\citenamefont {Dankert}\ \emph {et~al.}(2009)\citenamefont
  {Dankert}, \citenamefont {Cleve}, \citenamefont {Emerson},\ and\
  \citenamefont {Livine}}]{Dankert2009ca}%
  \BibitemOpen
  \bibfield  {author} {\bibinfo {author} {\bibfnamefont {C.}~\bibnamefont
  {Dankert}}, \bibinfo {author} {\bibfnamefont {R.}~\bibnamefont {Cleve}},
  \bibinfo {author} {\bibfnamefont {J.}~\bibnamefont {Emerson}}, \ and\
  \bibinfo {author} {\bibfnamefont {E.}~\bibnamefont {Livine}},\ }\href
  {\doibase 10.1103/PhysRevA.80.012304} {\bibfield  {journal} {\bibinfo
  {journal} {Phys. Rev. A}\ }\textbf {\bibinfo {volume} {80}} (\bibinfo {year}
  {2009}),\ 10.1103/PhysRevA.80.012304}\BibitemShut {NoStop}%
\bibitem [{\citenamefont {Rigetti}\ \emph {et~al.}(2012)\citenamefont
  {Rigetti}, \citenamefont {Gambetta}, \citenamefont {Poletto}, \citenamefont
  {Plourde}, \citenamefont {Chow}, \citenamefont {C\'orcoles}, \citenamefont
  {Smolin}, \citenamefont {Merkel}, \citenamefont {Rozen}, \citenamefont
  {Keefe}, \citenamefont {Rothwell}, \citenamefont {Ketchen},\ and\
  \citenamefont {Steffen}}]{rigetti2012}%
  \BibitemOpen
  \bibfield  {author} {\bibinfo {author} {\bibfnamefont {C.}~\bibnamefont
  {Rigetti}}, \bibinfo {author} {\bibfnamefont {J.~M.}\ \bibnamefont
  {Gambetta}}, \bibinfo {author} {\bibfnamefont {S.}~\bibnamefont {Poletto}},
  \bibinfo {author} {\bibfnamefont {B.~L.~T.}\ \bibnamefont {Plourde}},
  \bibinfo {author} {\bibfnamefont {J.~M.}\ \bibnamefont {Chow}}, \bibinfo
  {author} {\bibfnamefont {A.~D.}\ \bibnamefont {C\'orcoles}}, \bibinfo
  {author} {\bibfnamefont {J.~A.}\ \bibnamefont {Smolin}}, \bibinfo {author}
  {\bibfnamefont {S.~T.}\ \bibnamefont {Merkel}}, \bibinfo {author}
  {\bibfnamefont {J.~R.}\ \bibnamefont {Rozen}}, \bibinfo {author}
  {\bibfnamefont {G.~A.}\ \bibnamefont {Keefe}}, \bibinfo {author}
  {\bibfnamefont {M.~B.}\ \bibnamefont {Rothwell}}, \bibinfo {author}
  {\bibfnamefont {M.~B.}\ \bibnamefont {Ketchen}}, \ and\ \bibinfo {author}
  {\bibfnamefont {M.}~\bibnamefont {Steffen}},\ }\href {\doibase
  10.1103/PhysRevB.86.100506} {\bibfield  {journal} {\bibinfo  {journal} {Phys.
  Rev. B}\ }\textbf {\bibinfo {volume} {86}},\ \bibinfo {pages} {100506}
  (\bibinfo {year} {2012})}\BibitemShut {NoStop}%
\end{thebibliography}%

\end{document}